\documentclass{JFM-FLM_Au}

\usepackage{hyperref}

\newcommand{\I}{\mathrm{i}}
\newcommand{\D}{\mathrm{d}}
\newcommand{\E}{\mathrm{e}}

\renewcommand{\Re}{\mathbf {Re}}

\renewcommand{\vec}{\mathbf }

\lefttitle{A.\,O.~Korotkevich}
\righttitle{Journal of Fluid Mechanics}

\title{Weakly nonlinear interaction of capillary waves in a finite system: leading interaction process and scales' range of direct energy cascade}

\author{Alexander O. Korotkevich\aff{1,2}}

\affiliation{\aff{1}Center for Engineering Physics, Skolkovo Institute of Science and Technology, Bolshoy Boulevard, 30, bld. 1, Moscow, 121205 Russia
\aff{2}L.\,D.~Landau Institute for Theoretical Physics, RAS, prospekt Akademika Semenova, 1A, Chernogolovka, Moscow region, 142432 Russia}

\corresau{Alexander O. Korotkevich, \email{A.Korotkevich@Skoltech.ru}}

\begin{document}
\maketitle

\begin{abstract}
During comprehensive study of weakly nonlinear interaction of surface capillary waves,
processes of resonant and non-resonant interactions were considered both numerically and analytically: merging of two waves into one and waves on the ring (in Fourier space, isotropic spectrum) into larger diameter ring. It was shown numerically, that these resonant processes are the leading ones and other processes with respect to them are at least weaker if manifest themselves at all. It was confirmed, that resonant the processes are the major ones which contribute to the long time dynamics. In the case of isotropic turbulence of capillary waves the formation of wave turbulence's Zakharov-Filonenko spectrum is demonstrated. It was also shown, that this spectrum in finite systems has a finite range of scales. Due to finiteness of the numerical simulation or experimental area the discreteness of the wavenumbers grid arrest local in Fourier space resonant interaction when smaller scales are considered. Scaling of the range of realization of the Zakharov-Filonenko spectrum, depending on main parameters of the numerical or experimental setup (average steepness and characteristic size), is derived analytically and partially confirmed numerically.
\end{abstract}



\section{Introduction\label{sec:Intro}}
Capillary waves on a surface of deep water is a classical testbed for wave turbulence, due to almost the same equations as for gravity waves, which are important from the point of view of applications, e.g. wave forecasting. At the same time the system allows reasonable simplification (only three-wave processes are necessary from the point of view of physics) which improves computational time significantly. In a classical work~\citet{PZ1996} perhaps the first numerical confirmation of weakly turbulent universal~\citep{ZLF1992,Nazarenko2011} Kolmogorov-Zakharov (KZ) spectrum (specifically, Zakharov-Filonenko spectrum of direct cascade of energy for capillary waves, derived in~\citet{ZakharovPhD,ZF1967}) was demonstrated. Later series of works by~\citet{PY2014,PY2015,YulinPanPhD} determined the Kolmogorov constant for the spectrum, which was very significant support for the wave turbulence theory (WTT). Recently, there was a successful experimental study of capillary waves' turbulence in microgravity~\citet{BFMGF2020,FM2022}. One dimensional turbulence of capillary waves was studied experimentally and theoretically in recent papers~\citet{RF2021,KR2025}, where is was demonstrated that physics of nonlinear interactions changes dramatically.  More systematic review of this branch of nonlinear waves theory can be found in a recent paper~\citet{Galtier2021}.

Already at that of very first works, it become clear, that although exact resonant interactions, which is a fundamental process for WTT, never or practically never (depending on the physics in the system) happen due to exact resonance on a homogeneous grid of wavenumbers (standard setup for pseudo-spectral codes, exploiting fast Fourier transform), they still do realized due to broadening of the resonance because of nonlinearity~\citep{Pushkarev1999,PZ2000,CNP2001}. The first direct demonstration of a separate resonant process was presented in~\citet{DKZ2003cap} and extended to other types of processes in~\citet{KDZ2016}. It should be said that importance of discreteness of a wavenumbers grid for resonant processes even for analytical consideration was understood already by~\citet{KartashovaPhD,Kartashova1991} for Rossby waves and investigated further in~\citet{KNR2008}. Study of the influence of finite size effects on WTT was also active for gravity waves~\citep{ZKPD2005,KPRZ2008,ZP2022}, where in simulations with pumping discreteness of wavenumbers resulted in formation of so called condensate~\citep{Korotkevich2008PRL,Korotkevich2023}, which significantly changes the physics of wave interaction in the system~\citep{Korotkevich2012MCS,Korotkevich2013JETPL,KZ2015} or even results in a new spectrum appearance~\citep{KNPS2024}. In recent work~\citet{BBKKS2022} it was shown that too many resonant wavenumbers can influence the physics of the system as well, in accordance with~\citet{AS2006JFM}.

In this paper we perform comprehensive study of weakly nonlinear interaction of surface capillary waves, specifically separate resonant and non-resonant interaction processes on a homogeneous grid of wavenumbers. We considered the following fundamental for WTT processes both numerically and analytically: merging of two waves into one and waves on the ring (in Fourier space, isotropic spectrum) into larger diameter ring. It was shown numerically, that these resonant processes are the leading ones and give the major contribution after long times. In the case of isotropic turbulence of capillary waves the formation of wave turbulence's Zakharov-Filonenko spectrum is demonstrated. Unlike in the previous simulations we started from much lower values of wavenumbers, very narrow in scales and still obtained direct cascade spectrum formation. It was explained analytically that this configuration is not only feasible, but this particular KZ-spectrum for capillary waves in finite systems has a finite range of scales, thus starting from higher wavenumbers will not increase the dynamic range of scales. Due to finiteness of the numerical simulation or experimental area the discreteness of the wavenumbers grid effectively stops local in Fourier space transfer of energy flux due to resonant interactions for large wavenumbers. Analytical theory of the process is proposed and compared with the numerical simulations.

The plan of the paper is the following. We start from very brief problem formulation in Subsection~\ref{sec:problem_formulation}, with natural continuation by analytical investigation of decay and merge resonant processes on a grid of wavenumbers in Subsection~\ref{sec:resonant_conditions}. In the next Section~\ref{sec:numerical_results} we present numerical results. It starts with brief description of numerical setup in a subsection~\ref{sec:num_params}, and continues with simulation of the merging of asymmetric configuration of wave (Subsection~\ref{sec:merging}) which allows better separation of different processes and symmetric configuration of waves (Subsection~\ref{sec:merging_symm}) which is more relevant for isotropic spectra, popular in simulation of wave turbulence. Symmetric case is a building block for the case of an isotropic initial condition in a form of a ring, centered at the origin  of a Foruier or wavenumbers space. We start from easier to analyze ring of relatively large radius in Subsection~\ref{sec:ring43} and continue with more suitable for Kolmogorov-Zakharov spectrum generation case in Subsection~\ref{sec:ring5}, where substantial dynamic range of scales is available for formation of the Zakharov-Filonenko spectrum, which is observed and investigated. In the next Subsection~\ref{sec:break_ZF_spectrum} we explain how discreteness breaks Zakharov-Filonenko spectrum at large wavenumbers and derive some scalings of dynamic range with key parameters of the numerical setup. In the final Section~\ref{sec:conclusion} we summarize results of the work and present some future plans. In the Appendix~\ref{app:steepness} we moved a technical derivation of an average steepness dependence on parameters of a simple model of the spectrum for both cases of capillary and gravity waves, in dynamical equations and for wave kinetic equation (WKE).

\subsection{Problem formulation\label{sec:problem_formulation}}

Let us consider a potential flow of an ideal fluid of infinite depth with a free surface and use standard notations for velocity potential $\phi(\vec r, z, t),\vec r = (x,y); \vec v = \vec\nabla\phi$ and surface elevation $\eta(\vec r, t)$ with respect to an unperturbed state, when the fluid occupies $XY$-plane $z=0$.
\begin{figure}[tbh]
\centering
\includegraphics[width=10.0cm]{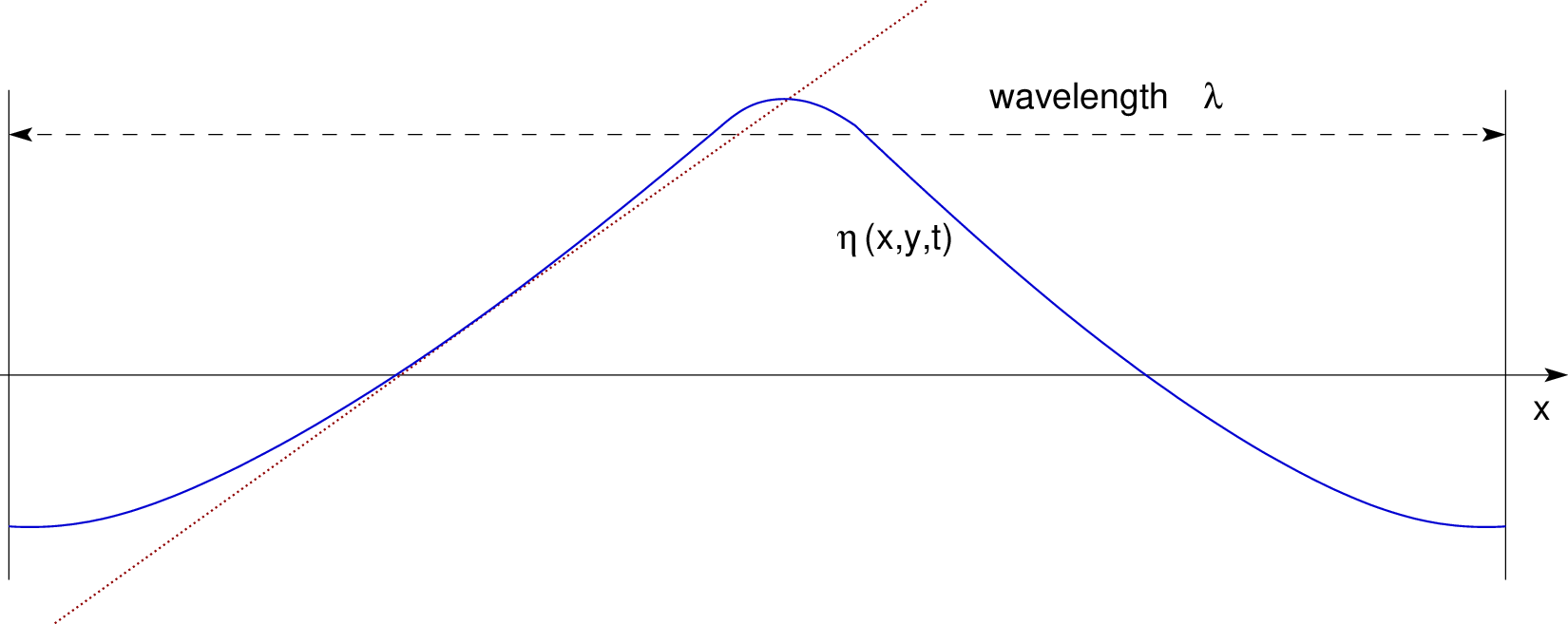}
\end{figure}
One of the most useful characteristics of the system is so called average steepness of the surface $\mu = \sqrt{\langle|\nabla\eta(\vec r,t)|^2\rangle}$ -- average slope of the surface. In many of interesting cases $\mu\simeq 0.1$ -- a small parameter.

Another approximation which is normally used is incompressibility of the fluid $(\vec\nabla \vec v) = \vec\nabla^2 \phi = 0$. The total energy of the system can be presented in the following form
\begin{equation}
H = \frac{\rho}{2} \int \D^2 r \int \limits_{-\infty}^{\eta} (\nabla \phi)^2 \D z +  \sigma \int (\sqrt{1 + (\nabla \eta)^2} - 1) \D^2 r,
\end{equation}
here $\sigma$ is the surface tension coefficient. Let us factor out $\rho$ and replace $\sigma$ with $\sigma'=\sigma/\rho$. After that without loss of generality fluid density can be put to $\rho=1$. Here and further, unless stated otherwise, integrals are considered over the whole fluid surface $\int \D^2 r=\int_{-\infty}^{+\infty} \D^2 r$.

It was shown by~\citet{Zakharov1967} that under these assumptions the fluid is a Hamiltonian system with respect to Hamiltonian variables $\eta$ and $\psi$
\begin{equation}
\label{Hamiltonian_equations}
\frac{\partial \eta}{\partial t} = \frac{\delta H}{\delta \psi}, \;\;\;\;
\frac{\partial \psi}{\partial t} = - \frac{\delta H}{\delta \eta},
\end{equation}
where $\psi = \phi (\vec r, z=\eta (\vec r,t), t)$ is a velocity potential on the surface of the fluid. In order to
calculate the value of $\psi$ we have to solve the Laplace equation in the domain with varying
surface $\eta$. One can simplify the situation, using the expansion of the Hamiltonian in powers of steepness $\mu$:
\begin{equation}
\label{Hamiltonian}
H = \frac{1}{2}\int\left( \sigma' |\nabla \eta|^2 + \psi \hat k  \psi \right) \D^2 r + \frac{1}{2}\int\eta\left[ |\nabla \psi|^2 - (\hat k \psi)^2 \right] \D^2 r,
\end{equation}
here $\hat k = \sqrt{-\vec\nabla^2}$.
Detailed derivation of this expansion for a more general case of gravity-capillary waves can be found in Supplemental materials of paper~\citet{Korotkevich2023}.

For Hamiltonian~\eqref{Hamiltonian} dynamical equations~\eqref{Hamiltonian_equations} acquire the following form
\begin{align}
\label{eta_psi_system}
\dot \eta &= \hat k  \psi - (\nabla (\eta \nabla \psi)) - \hat k  [\eta \hat k  \psi] - D_{\vec r},\\
\dot \psi &= \sigma' \vec\nabla^2 \eta - \frac{1}{2}\left[ (\nabla \psi)^2 - (\hat k \psi)^2 \right] - D_{\vec r} .\nonumber
\end{align}
Here $D_{\vec r}$ is some artificial damping term used to provide dissipation at small scales, modeling viscosity or used for stabilization of a numerical scheme. It has to be included symmetrically in both equations according to~\citet{DDZ2008}.

Taking into account weak nonlinearity, it is useful to consider Fourier transform of the $\eta$ and $\psi$:
$$
\psi_{\vec k} = \frac{1}{2\pi} \int \psi_{\vec r} \E^{\I {\vec k} {\vec r}} \D^2 r,\;\;
\eta_{\vec k} = \frac{1}{2\pi} \int \eta_{\vec r} \E^{\I {\vec k} {\vec r}} \D^2 r.
$$
Because $\psi(\vec r, t)$ and $\eta(\vec r, t)$ are real valued functions, they have to obey Hermitian symmetry: $\psi_{\vec k} = \psi_{-\vec k}^*,\;\eta_{\vec k} = \eta_{-\vec k}^*$, here ${}^*$ means complex conjugation. Even if one considers a plane wave propagating with a wave vector $\vec k$, spectra of both $\psi_{\vec k}$ and $\eta_{\vec k}$ will have peaks at both $\vec k$ and $-\vec k$, which is not convenient e.g. for visualization of the spectrum of the wave field. It is a standard approach~\citep{ZLF1992, Nazarenko2011} to introduce normal canonical variables:
$a_{\vec k}$ as shown below
\begin{equation}
\label{a_k_substitution}
a_{\vec k} = \sqrt \frac{\omega_k}{2k} \eta_{\vec k} + \I \sqrt \frac{k}{2\omega_k} \psi_{\vec k},
\end{equation}
here $\omega_k = \sqrt {\sigma' k^3}\sim k^{3/2}$ is the linear dispersion relation. In this variables Hamiltonian equations~\eqref{Hamiltonian_equations} take the form
\begin{equation}
\label{Hamiltonian_eqs_canonical}
\dot a_{\vec k} = -\I \frac{\delta H}{\delta a_{\vec k}^{*}}.
\end{equation}
The second Hamiltonian equation is the complex conjugated one. This is essentially representation of our wave field in terms of elementary excitations $a_{\vec k}$, which are nothing else but solutions of a linearized version of~\eqref{eta_psi_system} or plane waves.

\subsection{Resonant conditions\label{sec:resonant_conditions}}
If we are considering capillary waves and looking for the effects which will be present for so called "nonlinear times" (can be estimated by the order of magnitude as $1/\mu$ linear periods of a corresponding wave as show, for example, in the paper~\citet{ZKPD2005}), one needs to consider triads of waves which satisfy resonant conditions~\citep{ZLF1992}:
\begin{equation}
\label{resonant_conditions}
\Omega_{k_1 k_2}^{k} = \omega_{k_1} + \omega_{k_2} - \omega_{k} = 0,\;\;\;
\vec k_1 + \vec k_2 - \vec k = \vec 0.
\end{equation}
It can be shown that for dispersion relation for capillary waves $\omega_k\sim k^{3/2}$ on a homogeneous grid, typical for numerical simulation using pseudo-spectral codes, these conditions are never satisfied exactly in the grid nodes. Nevertheless, it was shown for the first time in~\citet{DKZ2003cap} and investigated in much more details in~\citet{KDZ2016} that, due to nonlinear broadening of the resonant curve, resonant conditions are replaced by weaker requirement
\begin{equation}
\label{nonlinear_resonant_conditions}
\Omega_{k_1 k_2}^{k} = \omega_{k_1} + \omega_{k_2} - \omega_{k} \lesssim \Gamma(\vec k,\mu,\ldots),\;\;\;
\vec k_1 + \vec k_2 - \vec k = \vec 0,
\end{equation}
which corresponds to appearance of a finite width for the infinitely thin curve~\eqref{resonant_conditions}, which can cover grid nodes. Here $\Gamma$ in~\eqref{nonlinear_resonant_conditions} represents this finite width of the resonant curve, which is different for different $\vec k$ and depends on the level of nonlinearity in the system, which can be estimated by $\mu$.

\begin{figure}[htb]
\centering
\includegraphics[width=0.9\textwidth]{"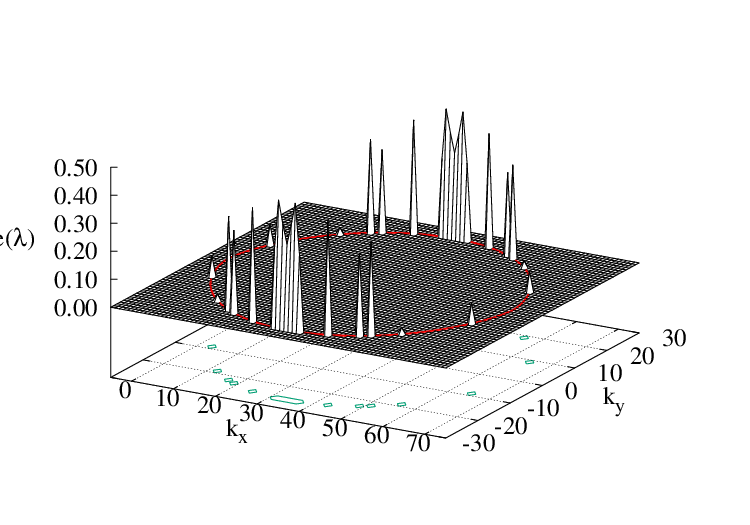"}
\caption{\label{fig:increment} Growth rate~\eqref{lambda_decay} as a function of $\vec k$ for decay of a monochromatic wave with the wave vector $\vec k_0=(68;0)$ on a grid, corresponding to average steepness $\mu=0.1$. Red line shows a resonant curve.}
\end{figure}

In previous works~\citep{DKZ2003cap,KDZ2016} the case of a decay of the monochromatic capillary wave in a periodic boundary conditions, thus, on a homogeneous grid of wavenumbers, was considered analytically and numerically. Specifically, initial wave with wavenumber $\vec k_0$ was decaying into two waves with wavenumbers $\vec k_1$ and $\vec k_2$ subject to resonant conditions~\eqref{nonlinear_resonant_conditions}. The exponential growth rate $\lambda$ for two waves $a_{\vec k_{1,2}}(t) = a_{\vec k_{1,2}}(0) \E^{\lambda t}\E^{\I\omega_{k_{1,2}} t}$, in the approximation of a constant amplitude $|a_{\vec k_0}|$ of a decaying wave (initial stage of decay, when amplitude of the initial wave is much larger than those of growing waves) is given by the following expression:
\begin{equation}
\label{lambda_decay}
\lambda =-\frac{\I}{2}\Omega_{k_1 k_2}^{k_0} +\sqrt{\left|\frac{2\pi}{L_x L_y} M_{\vec k_1 \vec k_2}^{\vec k_0} a_{\vec k_0}\right|^2 - \left(\frac{1}{2}\Omega_{k_1 k_2}^{k_0}\right)^2}.
\end{equation}
Here interaction coefficient $M_{\vec k_1 \vec k_2}^{\vec k_0}$ is defined in the following way:
\begin{align}
M_{\vec k_1 \vec k_2}^{\vec k_0} &= V_{\vec k_1 \vec k_2}^{k_0} - V_{-\vec k_0 \vec k_2}^{k_1} - V_{-\vec k_0 \vec k_1}^{k_2},\nonumber\\
V_{\vec k_1 \vec k_2}^{k_0} &= \sqrt {\frac{\omega_{k_1} \omega_{k_2} k_0}{8 k_1 k_2 \omega_{k0}}} L_{\vec k_1 \vec k_2},\label{matrix_element}\\
L_{\vec k_1 \vec k_2} &= (\vec k_1 \vec k_2) + |k_1||k_2|.\nonumber
\end{align}
Pay attention, that the constant $2\pi/(L_x L_y)$ depends on normalization of the Fourier series for $a_{\vec k_0}$. The surface $Re(\lambda(\vec k))$ for an average steepness $\mu=0.1$ and size of the periodic box $2\pi\times 2\pi$ (meaning that wavenumbers components are integers) is shown in Fig.~\ref{fig:increment}.

If one considers the similar growth rate for close to resonant merging of two waves with wavenumbers $\vec k_1$ and $\vec k_2$ into one with a wavenumber $\vec k_0$, instead of starting from the scratch it is easier to use the fact that time inversion changes the decay of a monochromatic capillary wave into merging of two waves into that wave. Also let us take into account the following relation. For three interacting waves (let us suppose for simplicity that they are in a perfect resonance) we have a total linear (by far the main contribution) energy:
\begin{equation}
\label{linear_energy_triad}
\omega_{k_0}|a_{\vec k_0}|^2 + \omega_{k_1}|a_{\vec k_1}|^2 + \omega_{k_2}|a_{\vec k_2}|^2 \approx const.
\end{equation}
Let us consider simplified case when $\omega_{k_1}=\omega_{k_2}$, then $\vec k_1$ and $\vec k_2$ are symmetric with respect to $\vec k_0$ and $a_{\vec k_{1,2}}(t)$ have the same growth rate $\lambda_1$, while the growth rate for $a_{\vec k_{0}}(t)$ can be denoted as $\lambda_0$. Then taking the time derivative of~\eqref{linear_energy_triad} and using ansatz $|a_{\vec k_i}(t)|^2\sim \E^{2\lambda_i t}$ one gets:
\begin{equation}
\label{derivative_linear_energy_triad}
\omega_{k_0}\lambda_0|a_{\vec k_0}|^2 + \omega_{k_1}\lambda_1 (|a_{\vec k_1}|^2 + |a_{\vec k_2}|^2) \approx 0.
\end{equation}
The resonant condition for frequencies $\omega_{k_0}=2\omega_{k_1}$ yields:
\begin{equation}
\label{lambdas_relation}
\lambda_0|a_{\vec k_0}|^2 \approx -\lambda_1 \frac{|a_{\vec k_1}|^2 + |a_{\vec k_2}|^2}{2}.
\end{equation}
Noting in~\eqref{lambdas_relation} that growth for $a_{\vec k_0}$ means decay for $a_{\vec k_{1,2}}$ and vise versa and using time reversal argument mentioned before, for the the process of merging $a_{\vec k_1}$ and $a_{\vec k_2}$ into $a_{\vec k_0}$ the exponential growth rate is:
\begin{equation}
\label{lambda_merge}
\lambda_0 =-\frac{\I}{2}\Omega_{k_1 k_2}^{k_0} +\sqrt{\left[\frac{2\pi}{L_x L_y} M_{\vec k_1 \vec k_2}^{\vec k_0}\right]^2 \frac{|a_{\vec k_1}|^2 + |a_{\vec k_2}|^2}{2} - \left(\frac{1}{2}\Omega_{k_1 k_2}^{k_0}\right)^2}.
\end{equation}

\section{Numerical results\label{sec:numerical_results}}
In this section we are going to use numerical simulation in the framework of the system~\eqref{eta_psi_system} to figure out what nonlinear processes are producing lasting effect on the system and what specific processes are present at all. The main goal is to understand what nonlinear processes are responsible for formation of direct cascade of wave turbulence of capillary waves and what particular scales are involved in interaction. In order to do it, we start with relatively easy to analyze merging process of two asymmetric waves, then proceed to similar symmetric case, and then continue to two cases of isotropic decaying turbulence: with relatively small initial scales (large $k$) and large scale (small $k$) initial excitation. Relatively simple initial cases of two initial plane waves allow to analyze practically every available process, while more complex interactions in the case of isotropic ensembles of waves are closer to realistic wave turbulence numerical experiments. All numerical simulations implement so called decaying turbulence, similar to~\citet{Onorato2002,ZKPD2005,ZKPR2007,KPRZ2008,PY2015}

\subsection{Numerical scheme parameters\label{sec:num_params}}
Let us specify the damping term in the dynamical equations~\eqref{eta_psi_system} in accordance with~\cite{DDZ2008}:
\begin{align}
\dot \eta &= \ldots - F^{-1}[\gamma_k \eta_{\vec k}],\;\;
\dot \psi = \ldots - F^{-1}[\gamma_k \psi_{\vec k}],\nonumber\\
D_{\vec k} &= \gamma_k \psi_{\vec k},\;\;\gamma_{k} = \gamma_0 (k - k_d)^2, \; k > k_d. 
\end{align}
For $k\le k_d$ there is no artificial dissipation $\gamma_{k}=0$. Parameter $\gamma_0$ is chosen automatically to be the minimal one ensuring difference between the largest harmonic and the smallest (furthers) one to be at least $10^6$. Times step adjustment procedure as well as accuracy control approach are described in~\cite{KDZ2016}.
Damping starts at $k_d = 2/3 k_{max}$. Unless directly stated otherwise, we use grid resolutions $N_x=N_y=512$, $k_{max}=256$ and, correspondingly, $k_d=170$.
Simulation region is $L_x = L_y = 2\pi$ with periodic boundary conditions in both directions. In this case all wavenumbers projections $k_x$ and $k_y$ are integer numbers. In all simulations $\sigma'=1$ for simplicity and time measured in periods of one of relevant waves.

In all of the cases we consider decaying turbulence, so no artificial pumping terms are present. Initial condition in wavenumbers space have random phases for  all harmonics, specifically phase of every harmonic is a uniformly distributed random number from an interval $[0,2\pi)$. With exception of initial excitations, all initial harmonics have the same amplitude $10^{-12}$, mimicking background noise. Numerical scheme used for integration is described in details in~\citet{KDZ2016} and called Hamiltonian integration. The same scheme, originally proposed in~\citet{DNPZ1992}, can be efficiently applied to other Hamiltonian systems as well~\citep{SDKL2021}.

\subsection{Merging of two waves: asymmetric case\label{sec:merging}}
\begin{figure}[htb]
\centering
\includegraphics[width=0.45\textwidth]{"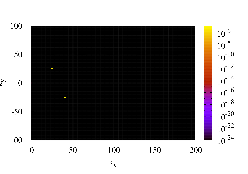"}
\includegraphics[width=0.45\textwidth]{"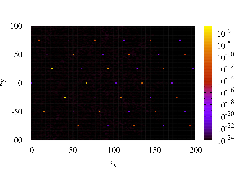"}
\caption{\label{fig:asymm_initial}Surface $|a_{\vec k}|^2$. (Left) Initial condition for Subsection~\ref{sec:merging}. (Right) The same surface for the moment of time $t=17.8 T_{68}$.}
\end{figure}

As a case which is simplest to analyze, we consider initial condition of just two asymmetrically placed plain waves. In a sense this experiment is an inverse of one from a paper~\citet{DKZ2003cap}. We start with two plain waves $\vec k_1 = (26;25)$ and $\vec k_2 = (42;-25)$ which are close to exact resonance~\eqref{resonant_conditions} according to which they suppose to merge into the wave $\vec k_0=(68;0)$. Amplitudes of these initial waves are equal and chosen to give average steepness $\mu\approx 0.1$. What is important, they are also relatively close to the maxima of interaction coefficient $M_{\vec k_1 \vec k_2}^{\vec k_0}$, which are located at positions $(34;26)$ and $(34;-26)$, as can it be seen from Fig.~\ref{fig:increment}. In this subsection we deliberately break the symmetry of initial excitations in order to make it easier to separate different nonlinear processes. Initial condition is represented in the left panel of Fig.~\ref{fig:asymm_initial}.
It is reasonable to measure time in some characteristic times. Here we use period $T_{68}$ of the wave with wavenumber $\vec k_0=(68;0)$ which has to appear as a result of merging of two initial waves. Nonlinear time can be roughly estimated as $T_{68}/\mu \approx 10 T_{68}$. Right panel of Fig.~\ref{fig:asymm_initial} represents result of waves dynamics after $17.8 T_{68}$ as a surface $|a_{\vec k}|^2$. One can see set of separate plane waves which appeared as a result of nonlinear interactions. It is important to say that here we can observe results of all nonlinear interactions, not only resonant ones. In other words, condition on wavenumbers in~\eqref{resonant_conditions} has to be satisfied (this is just a result of Fourier transform of a nonlinear term, which generates factor similar to $\delta(\vec k_1 + \vec k_2 - \vec k_0)$), while frequency condition will affect the situation on longer times. In order to understand nearly every generated plane wave, we demonstrate series of figures with specific interactions clearly illustrated. 

\begin{figure}[htb]
\centering
\includegraphics[width=0.45\textwidth]{"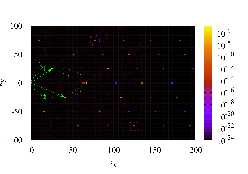"}
\includegraphics[width=0.45\textwidth]{"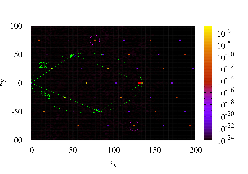"}
\caption{\label{fig:asymm_merging_initial}Surface $|a_{\vec k}|^2$. (Left) Merging of two initial asymmetric waves into the third one on a resonant curve. (Right) Merging of second harmonics of original waves into the wave on a resonant curve. Second and third harmonics of initial waves are shown by blue and magenta circles correspondingly. Moment of time $t=17.8 T_{68}$.}
\end{figure}
In the left panel of Fig.~\ref{fig:asymm_merging_initial} we clearly see merging process $\vec k_1 + \vec k_2$ for initial waves (shown by green vectors), generating the one with wavenumber $\vec k_0=(68;0)$ (shown by red vector), which we expect to survive on long times, as it satisfies resonant conditions (resonant manifold is shown by dashed line). At the same time, because second harmonics (shown by blue circles) of initial waves (shown by green vectors) are also generated, they merge into the wave with wavenumber $2\vec k_1 + 2\vec k_2=2\vec k_0=(2\times 68;0)$ (shown by red vector). This is shown in the right panel of a Fig.~\ref{fig:asymm_merging_initial}, where the corresponding resonant manifold for second harmonics is shown by dashed line. Similar processes with third harmonics generate more waves visible along a horizontal line $k_y=0$.

\begin{figure}[htb]
\centering
\includegraphics[width=0.45\textwidth]{"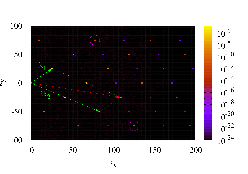"}
\includegraphics[width=0.45\textwidth]{"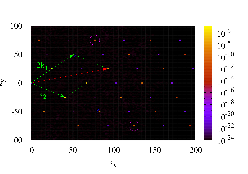"}
\caption{\label{fig:asymm_merging_trinary}Surface $|a_{\vec k}|^2$. Merging of a second harmonics of one of the initial asymmetric waves with another initial wave into the third one. Second and third harmonics of initial waves are shown by blue and magenta circles correspondingly. Moment of time $t=17.8 T_{68}$.}
\end{figure}
Due to interaction of second harmonics (shown by blue circles starting from Fig.~\ref{fig:asymm_merging_initial}) and initial harmonics we get extra harmonics generation. These processes are shown in Fig.~\ref{fig:asymm_merging_trinary}. In the left panel one can see merging of waves corresponding to the process $2\vec k_1 + \vec k_2$ (result is shown by the red vector), while on the right panel the process $\vec k_1 + 2\vec k_2$ generates a new wave, shown by the red vector. Similar processes involving third harmonics (shown by magenta circles starting from Fig.~\ref{fig:asymm_merging_initial}) generate other waves visible along straight line connecting $\vec k_2$ and $2\vec k_1 + \vec k_2$. The same description is true for waves generated along the line $\vec k_1$ and $\vec k_1 + 2\vec k_2$.

\begin{figure}[htb]
\centering
\includegraphics[width=0.45\textwidth]{"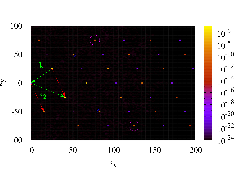"}
\includegraphics[width=0.45\textwidth]{"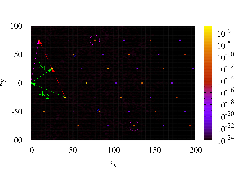"}
\caption{\label{fig:asymm_merging_less_trivial}Surface $|a_{\vec k}|^2$. Merging of a second harmonic of one of the initial waves with another initial wave into the third one. Second and third harmonics of initial waves are shown by blue and magenta circles correspondingly. Moment of time $t=17.8 T_{68}$.}
\end{figure}
Less trivial processes generate all other waves visible in the right panel of Fig.~\ref{fig:asymm_initial}. In the Fig.~\ref{fig:asymm_merging_less_trivial} one can see examples of a couple of such processes. In the left panel of Fig.~\ref{fig:asymm_merging_less_trivial} we see generation of a wave $-\vec k_1 + \vec k_2$, while in the right panel of the same Figure one can see the process $2\vec k_1 - \vec k_2 = \vec k_1 + (\vec k_1 - \vec k_2)$ resulting in a new wave.

\begin{figure}[htb]
\centering
\includegraphics[width=0.45\textwidth]{"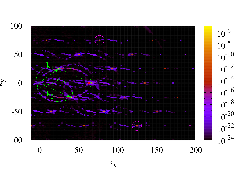"}
\includegraphics[width=0.45\textwidth]{"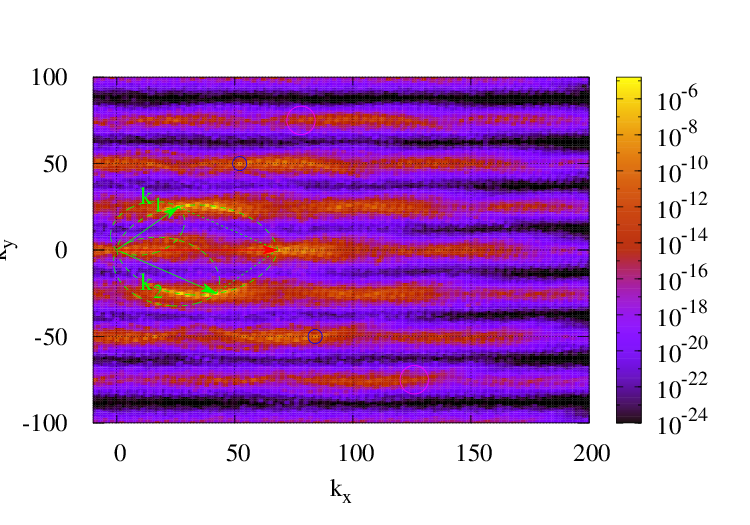"}
\caption{\label{fig:asymm_merging_longer_times}Surface $|a_{\vec k}|^2$. Merging of initial asymmetric waves after significant time. (Left) Moment of time $t=892 T_{68}$. (Right) Moment of time $t=1785 T_{68}$. Second and third harmonics of initial waves are shown by blue and magenta circles correspondingly.}
\end{figure}
At longer times, after $T_{68}/\mu^2 \approx 100 T_{68}$, one should expect manifestation of both resonant interactions~\eqref{resonant_conditions}, including condition on frequencies. In Fig.~\ref{fig:asymm_merging_longer_times}, left panel shows the plane of harmonics in the moment of time $t=892 T_{68}$. The reason for such a long time is that we would like to see resonant processes manifested for most of the relevant scales. For example, $|\vec k_1|\approx 38$, which means that $T_{38} = (68/38)^{3/2} T_{68} \approx 2.6 T_{68}$. The maximum of a growth rate for decay of a monochromatic capillary wave is close to the middle of the corresponding resonant curve~\citep{KDZ2016}, thus characteristic period is $7.3 T_{68}$, which results in estimation for times when resonant interactions will play a major role for decay of a plane wave with a wavenumber $\vec k_1$ of the order of $7.3\times 100 T_{68}$. On the right panel of a Fig.~\ref{fig:asymm_merging_longer_times} one can clearly see several resonant processes. First of all, we see decay of individual initial plane waves $\vec k_1$ and $\vec k_2$ as it was shown previously for one plane wave in~\citet{DKZ2003cap}. Another process is decay of a wave $\vec k_0=(68;0)$, into which initial waves merged and which at time $t=892 T_{68}$ has long achieved an amplitude comparable with amplitudes of initial waves. Resonant manifolds (oval curves) for these three processes are shown by green dashed lines. Also results of translation of these curves by different wave vectors, shown in the left panel of Fig.~\ref{fig:asymm_merging_initial}, are visible as well, together with weaker amplitudes corresponding to decay of the wave $\vec 2k_0=(136;0)$. For twice larger time $t=1785 T_{68} \approx 2\times 892 T_{68}$ all these processes are manifested even more and interaction processes blur most of the secondary processes, while decay and merging of main triad become more visible. 

\begin{figure}[htb]
\centering
\includegraphics[width=0.45\textwidth]{"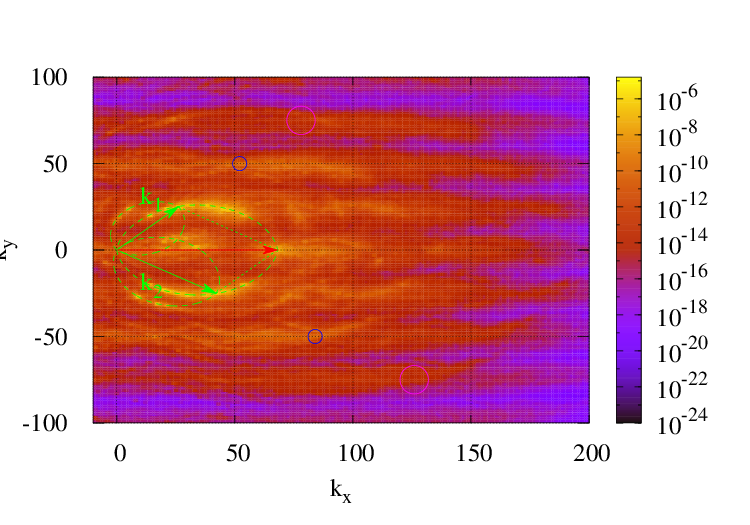"}
\includegraphics[width=0.45\textwidth]{"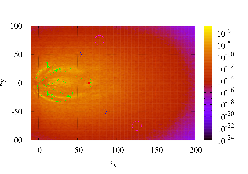"}
\caption{\label{fig:asymm_merging_longest_times}Surface $|a_{\vec k}|^2$. Merging of initial asymmetric waves after very long time. (Left) Moment of time $t=2677 T_{68}$. (Right) Moment of time $t=74092 T_{68}$. Second and third harmonics of initial waves are shown by blue and magenta circles correspondingly.}
\end{figure}
At the time $t=2677 T_{68} \approx 3\times 892 T_{68}$ the blurring of the secondary processes is even stronger, while main resonant processes become even more visible. The key feature of this picture is absence of separate multiple harmonics. Leftovers of nonresonant processes become less and less distinguishable. This moment of time is shown in the left panel of Fig.~\ref{fig:asymm_merging_longest_times}. Finally, at the longest time of simulation $t=74092 T_{68}$ we see more or less homogeneous sea of harmonics with strong peaks close to the resonant curves of resonant decay and merging of the the initial triad $\vec k_0$, $\vec k_1$ and $\vec k_2$. All other processes are significantly weaker and blurred to the point when individual plane waves become indistinguishable from each other by amplitude. It should be emphasized that all these pictures are in logarithmic scale, so relatively small difference in color can represent few orders of magnitude difference.

\begin{figure}[htb]
\centering
\includegraphics[width=0.9\textwidth]{"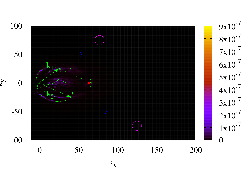"}
\caption{\label{fig:asymm_merging_longest_time_linear} Wave field in Fourier space after Linear scale. $t=74092 T_{68}$.}
\end{figure}
The same picture but in linear scale is represented in Fig.~\ref{fig:asymm_merging_longest_time_linear}. One can see that indeed, only main resonant processes remain. Specifically, the initial waves $\vec k_1$, $\vec k_2$ (shown by the green vectors) and the wave $\vec k_0$ in which they merged (shown by the red vector) are the brightest ones. Other waves are visible only along the corresponding decay curves for every individual wave of the main triad. None of the nonresonant processes, e.g. multiple harmonics generation, are distinguishable. This confirms that the resonant processes are indeed the ones which provide the major contribution in a long run.

\subsection{Merging of two waves: symmetric case\label{sec:merging_symm}}
\begin{figure}[htb]
\centering
\includegraphics[width=0.45\textwidth]{"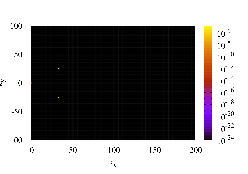"}
\includegraphics[width=0.45\textwidth]{"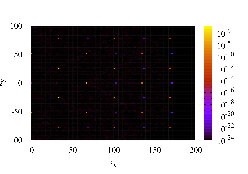"}
\caption{\label{fig:symm_initial}Surface $|a_{\vec k}|^2$. (Left) Initial condition for Subsection~\ref{sec:merging_symm}. (Right) The same surface for the moment of time $t=17.8 T_{68}$.}
\end{figure}

Now let us consider initial condition of two symmetrically placed plain waves, which is closer to standard numerical experiment simulating forced turbulence of capillary waves. For example, in both major examples of such works~\citet{PZ1996,PY2014}, isotropic pumping was used. If one would have a look at Fig.~\ref{fig:increment} and compare formulae~\eqref{lambda_decay} and~\eqref{lambda_merge}, it is clear that the most efficient merging of two waves with wavenumbers $\vec k_1$ and $\vec k_2$ into $\vec k_0$ will be observed if these wave are positioned symmetrically, close to the maxima of interaction coefficient $M_{\vec k_1 \vec k_2}^{\vec k_0}$. We shall consider this case in details as it is very important for following subsections. Here we consider two symmetric plain waves $\vec k_1 = (34;25)$ and $\vec k_2 = (34;-25)$ which are close to exact resonance~\eqref{resonant_conditions} according to which they suppose to merge into the wave $\vec k_0=(68;0)$. Amplitudes of these initial waves are equal and chosen to give average steepness $\mu\approx 0.1$. What is important, they are exactly at the maxima of interaction coefficient $M_{\vec k_1 \vec k_2}^{\vec k_0}$, as it can be seen from Fig.~\ref{fig:increment}. Initial condition is represented in the left panel of Fig.~\ref{fig:symm_initial}.
Once again we measure time in periods $T_{68}$ of the wave with wavenumber $\vec k_0=(68;0)$ which has to appear as a result of merging of two initial waves. Right panel of Fig.~\ref{fig:symm_initial} represents result of waves dynamics after $17.8 T_{68}$, meaning around one nonlinear time $\sim 10 T_{68}$, as a surface $|a_{\vec k}|^2$. One can see slightly different set of separate plane waves, compared with the right panel of Fig.~\ref{fig:asymm_initial}, which appeared as a result of nonlinear interactions. Once again we can observe results of all nonlinear interactions, not only resonant ones.

\begin{figure}[htb]
\centering
\includegraphics[width=0.45\textwidth]{"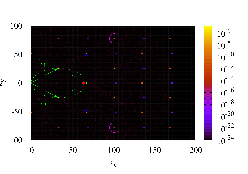"}
\includegraphics[width=0.45\textwidth]{"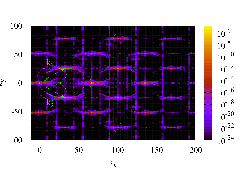"}
\caption{\label{fig:symm_merging_initial_later}Surface $|a_{\vec k}|^2$. (Left) Merging of two initial symmetric waves into the third one on a resonant curve. Moment of time $t=17.8 T_{68}$. (Right) The same process after significant time. Moment of time $t=892 T_{68}$. Second and third harmonics of initial waves are shown by blue and magenta circles correspondingly.}
\end{figure}
The main merging process of two initial symmetric waves into the third one $\vec k_1 + \vec k_2 = \vec k_0$ is shown in the left panel of Fig.~\ref{fig:symm_merging_initial_later}. One can see that initial wave are chosen on the resonant curve corresponding to the decay of the plain wave wave with wavenumber $\vec k_0$, shown with green dashed line. As in the previous Subsection~\ref{sec:merging} for asymmetric waves, after significant time $t=892 T_{68}$ we see formation of a decay of a wave corresponding to $\vec k_0$ along the resonant curve with translation of the decay ``oval'' over the whole wavenumbers plane by every of the waves, which were present in the right panel of Fig.~\ref{fig:symm_initial}.

\begin{figure}[htb]
\centering
\includegraphics[width=0.45\textwidth]{"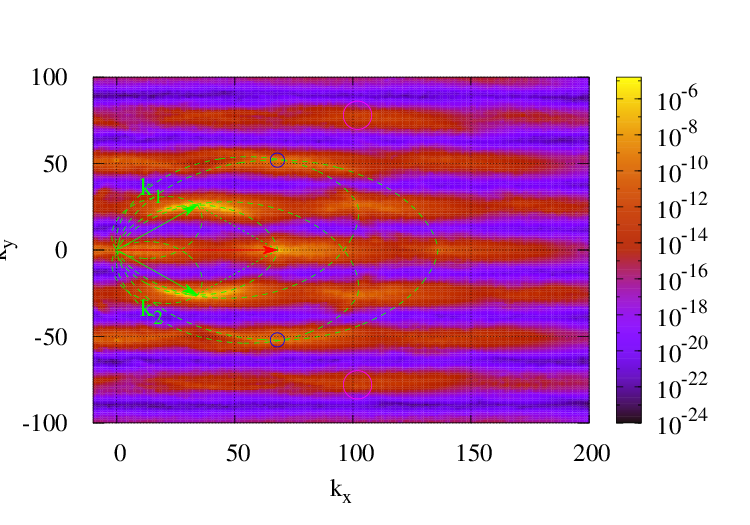"}
\includegraphics[width=0.45\textwidth]{"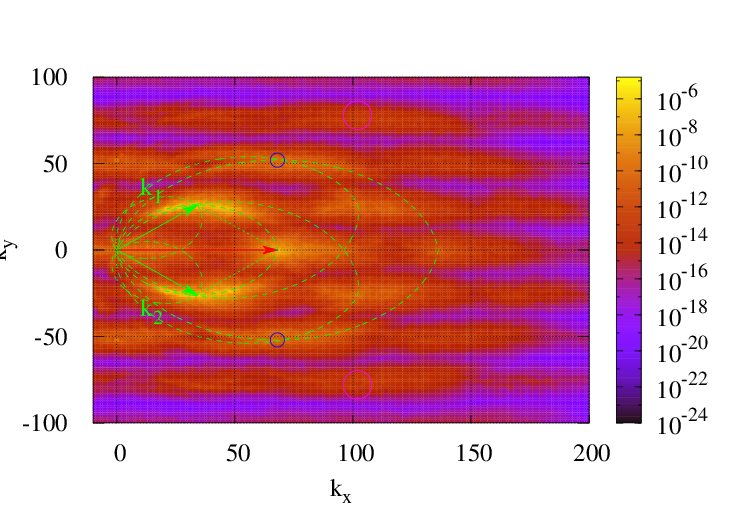"}
\caption{\label{fig:symm_merging_longer_times}Surface $|a_{\vec k}|^2$. Merging of initial symmetric waves after long time. (Left) Moment of time $t=1785 T_{68}$. (Right) Moment of time $t=2356 T_{68}$. Second and third harmonics of initial waves are shown by blue and magenta circles correspondingly.}
\end{figure}
Like in the right panel of Fig.~\ref{fig:asymm_merging_longer_times} and the left panel of Fig.~\ref{fig:asymm_merging_longest_times} for asymmetric initial condition, in the Fig.~\ref{fig:symm_merging_longer_times} we observe gradual blurring of nonresonant processes and only resonant processes of decay of initial waves $\vec k_1$, $\vec k_2$ and the result of their merging $\vec k_0 = \vec k_1 + \vec k_2$ remain visible. It is interesting to note, that surfaces $|a_{\vec k}|^2$ in the left panel of Fig.~\ref{fig:asymm_merging_longest_times} and the right panel of Fig.\ref{fig:symm_merging_longer_times} become very similar, as they are dominated by interactions of waves on the same resonant curve for decay of the monochromatic wave with wavenumber $\vec k_0$, which is the result of merging of initial wave in both cases.

\subsection{Ring in $\vec k$-space of a central radius $43$\label{sec:ring43}}
\begin{figure}[htb]
\centering
\includegraphics[width=0.45\textwidth]{"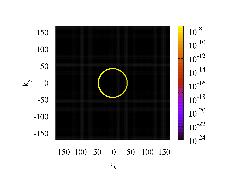"}
\includegraphics[width=0.45\textwidth]{"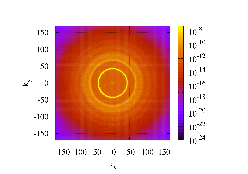"}
\caption{\label{fig:ring43_initial}Surface $|a_{\vec k}|^2$. (Left) Initial condition for Subsection~\ref{sec:ring43}. (Right) The same surface for the moment of time $t=17.8 T_{68}$.}
\end{figure}

Now let us consider the case which is a standard for numerical simulation of forced waves turbulence on the 2D surface of 3D fluid~\citep{PZ1996,DKZ2003grav,DKZ2004,Korotkevich2008PRL,Korotkevich2012MCS,Korotkevich2013JETPL,PY2014,PY2015,Korotkevich2023}. In all of these works the forcing is isotropic with respect to an angle and concentrated in some relatively narrow range of scales. Essentially, in wavenumbers space pumping force located on a ring with random phase of every harmonic. Due to such a configuration, averaging of the resulting spectra can be considered as a cheap replacement for an averaging over realizations, which is necessary for comparison with solutions of a wave kinetic equation~\citep{ZLF1992,Nazarenko2011}. So let us consider initial condition in a form of a narrow ring in $\vec k$-space. As before, we shall start from a case which allows us to separate different interaction processes and use a correspondence with a previous results. The ring will cover vectors $\vec k_1 = (34;25)$ and $\vec k_2 = (34;-25)$ from the previous Subsection~\ref{sec:merging_symm}, each of which has magnitude $|\vec k_1|=|\vec k_2|\approx 42.2$. The internal radius of the ring is $42$ and the external radius is $44$, the amplitudes of harmonics of the ring are the same and chosen to obtain average steepness $\mu=0.1$. Harmonics outside of the ring had an amplitude $10^{-12}$, simulating noise. Phase of every harmonic was a uniformly distributed random number from an interval $[0,2\pi)$. Surface $|a_{\vec k}|^2$ of the described initial condition is shown in a left panel of Fig.~\ref{fig:ring43_initial}.
After the same time as for the picture on the left panel of Fig.~\ref{fig:symm_merging_initial_later} we observe generation of new harmonics over the whole plane of wavenumbers as it is shown in the right panel of Fig.~\ref{fig:ring43_initial}. It is clear, that system of rings is formed. Using analogy with the dynamics of the two symmetrically placed initial waves, one could think that these rings are multiple harmonics of the initial ring, which is wrong. Every ring here is a result of a resonant interaction, meaning that it will persist for long times.
\begin{figure}[htb]
\centering
\includegraphics[width=0.45\textwidth]{"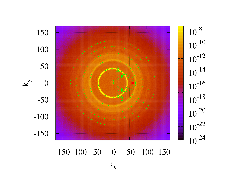"}
\includegraphics[width=0.45\textwidth]{"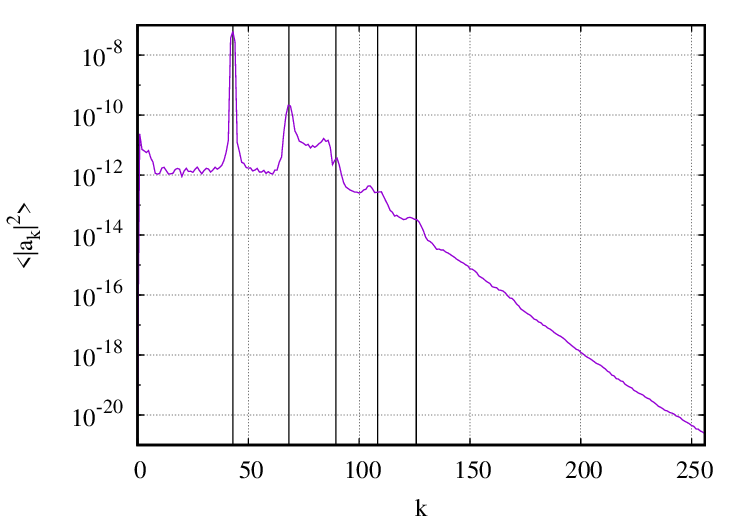"}
\caption{\label{fig:ring43_initial_later}Surface $|a_{\vec k}|^2$. (Left) Result of simulation for the moment of time $t=17.8 T_{68}$ with the resonant curve corresponding to the resonant merging of two waves from the initial ring. (Right) The same surface after averaging over angle. Thick vertical lines show positions of resonant merging of two waves: one on the initial ring and another on the same or different ring.}
\end{figure}
In order to understand specific interaction resulting in these rings formation, let us have a look at the left panel of Fig.~\ref{fig:ring43_initial_later}. Here we reproduced the same resonant process as we already considered during merging of symmetric waves, shown in the left panel of Fig.~\ref{fig:symm_merging_initial_later}. Indeed, for every direction one can choose symmetrically two waves on the ring (e.g. like $\vec k_1 = (34;25)$ and $\vec k_2 = (34;-25)$), which will merge into the third wave with a wave vector $\vec k_0$ of the length $|\vec k_1|^{3/2}+|\vec k_2|^{3/2} = |\vec k_0|^{3/2}$ according to resonant conditions~\eqref{resonant_conditions}. If we choose two waves on the initial ring with the same length of the wave vector $|\vec k_1|$, then $|\vec k_0|= 2^{2/3} |\vec k_1|\approx 1.6 |\vec k_1|$. The middle of the initial ring corresponds to $k=43$, so the ring produced by merging of two waves has to have radius $2^{2/3} 43\approx 69$. And indeed this is the central radius of the next ring, as it can be seen in Fig.~\ref{fig:ring43_initial_later} (see the dashed green circle of the radius slightly larger, that the initial ring). The most convenient way to track positions of the rings is to use angle averaged spectrum, which is shown in the right panel of Fig.~\ref{fig:ring43_initial_later}. Thick vertical lines show initial ring and further rings according to the resonant interactions (specific origin of some of them will be revealed in the next paragraph).

If one would apply the same logic to the new ring of the central radius $k=69$ (second smallest in radius in the left panel of Fig.~\ref{fig:ring43_initial_later}), we will get the ring of a radius $2^{2/3} 69\approx 110$, which is indeed visible. At the same time, there is a ring of an intermediate radius as well. In order to understand its origin, we need to return to Subsection~\ref{sec:merging}. We can consider two waves of different radii: $k_i = 43$ (initial ring) and $k_2 = 69\approx (2 k_i^{3/2})^{2/3}$ (results of merging of two waves in the initial ring). In the results of resonant merging, according to the resonant conditions~\eqref{resonant_conditions} we are supposed to get a wave of a radius $(k_i^{3/2} + k_2^{3/2})^{2/3}=(3 k_i^{3/2})^{2/3}\approx 90$, which exactly corresponds to the desired ring of the intermediate radius. This interaction is strong, because the magnitude of any wave in the initial ring is several orders of magnitude larger, than the one for any wave in further rings. Following Eq.~\eqref{lambda_merge}, this amplitude will dominate completely the one from the larger ring. Simple continuation of the same analysis for further rings shows, that the radii for all rings will be:
\begin{equation}
\label{rings_positions}
k_n = (n k_i^{3/2})^{2/3}=n^{2/3} k_i.
\end{equation}
These are exactly the positions of the thick vertical lines in the right panel of Fig.~\ref{fig:ring43_initial_later}.

\begin{figure}[htb]
\centering
\includegraphics[width=0.45\textwidth]{"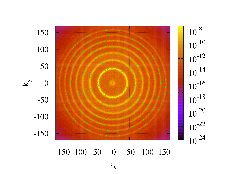"}
\includegraphics[width=0.45\textwidth]{"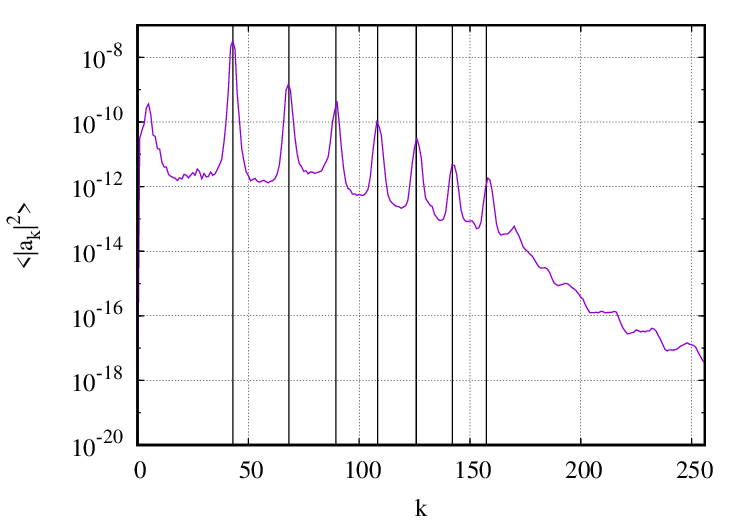"}
\caption{\label{fig:ring43_latest}Surface $|a_{\vec k}|^2$. (Left) Result of simulation for the moment of time $t=1785 T_{68}$ with the resonant curve corresponding to the resonant merging of two waves from the initial ring. Radii of the green dashed rings are given by Eq.~\eqref{rings_positions}. (Right) The same surface after averaging over angle. Thick vertical lines show positions of resonant merging of two waves: one on the initial ring and another on the same or different ring. Positions of the lines correspond to the Eq.~\ref{rings_positions}.}
\end{figure}
As we already have seen in Subsections~\ref{sec:merging}-\ref{sec:merging_symm}, after dynamics over a significant time we are left only with resonant interaction. This is exactly the case shown in Fig.~\ref{fig:ring43_latest} which corresponds to approximately 100 times further moment of time, than in Fig.~\ref{fig:ring43_initial_later}. Radii of the green dashed rings in the left panel and positions of thick vertical lines in the right panel are given by Eq.~\eqref{rings_positions}. It is clear, that the resonant process of merging of waves in different rings with the waves in the strongest initial ring describes the whole picture in all details.

The chosen relatively large radius of initial ring $k_i=43$ allowed us to both use direct comparison with the results of simulations in Subsection~\ref{sec:merging_symm} and to have the resonant rings well separated, which is useful for analysis. At the same time it is clear, that if we consider merging of a wave from the initial ring (smallest radius) with the wave from some large radius we will have two waves with wavenumbers very far from the maxima of the interaction coefficient, as can be seen from Fig.~\ref{fig:increment}. So eventually, even the large amplitude of the initial ring will be overpowered by the smallness of the interaction coefficient. As a result, again merging of two waves on the same ring, which corresponds to the maxima of the interaction coefficient, will become the major process. Unfortunately, radius of already the third such ring $(2)^{2/3}110\approx 176$ (let us recall, that we followed this process from merging of two waves in the initial ring $(2)^{2/3}43\approx 69$ to the merging of two waves in the resulting ring $(2)^{2/3}69\approx 110$) is beyond the dissipative wavelength $k_d=170$ and is strongly suppressed. So one needs to consider initial ring of significantly smaller radius, similar to the one which was used in previous simulations of forced turbulence of capillary waves.

While relatively large $k_i$ allows to distinguish different nonlinear processes, the broadening of the rings although happening, is way smaller, than the distance between them. This was convenient to our purposes of studying the separate interaction processes but prevents formation of the weakly turbulent Kolmogorov-Zakharov spectrum. This is another reason to consider smaller initial ring.

\subsection{Ring in $\vec k$-space of a central radius $5$\label{sec:ring5}}
Because of the reasons discussed in the end of the previous Subection~\ref{sec:ring43}, as initial condition this time we consider small ring with the central radius $k_i=5$ and the same width $3$. In other words, harmonics $4\le k \le 6$ have constant amplitude, giving the average steepness $\mu=0.2$. Because we are not ``backward compatible'' anymore (not using any similar scales from the previous examples), we will express time in $T_{5}=2\pi/\omega_{k_i}$ -- periods of the wave with wavenumber $k_i=5$. This is the only relevant timescale in the system. After time $t=90T_{5}$, in Fig.~\ref{} one can observe formation of the spectrum which looks like the decaying power-like function. In order to determine the slope we use a standard representation in double logarithmic scale (logarithm of both $X$- and $Y$-axis). Once again we use thick vertical lines to mark positions according to~\eqref{rings_positions}. Because due to broadening of the peaks only few first ones are visible, we limit ourselves with few of such marks. Nevertheless, while we can see the peaks they do coincide with predictions perfectly.

\begin{figure}[htb]
\centering
\includegraphics[width=0.45\textwidth]{"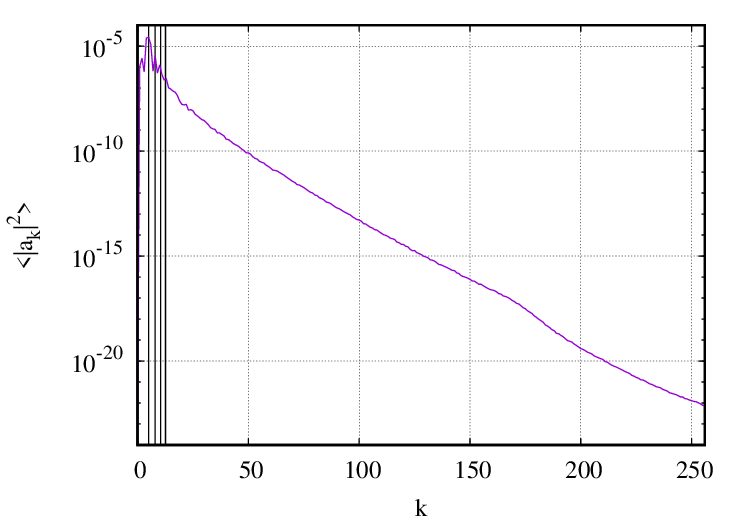"}
\includegraphics[width=0.45\textwidth]{"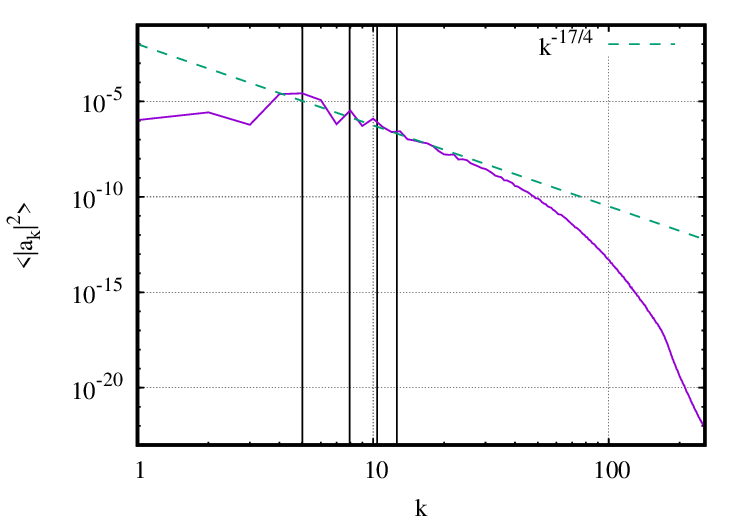"}
\caption{\label{fig:ring5_short}Angle averaged $|a_{\vec k}|^2$ at the moment of time $t=90 T_{5}$. (Left) linear scale on $k_x$-axis. (Right) The same surface in logarithmic scale on $k_x$-axis. One can notice formation of powerlike spectrum in significant range of scales. Green line corresponds to weakly turbulent Zakharov-Filonenko spectrum $|a_{\vec k}|^2\sim k^{-17/4}$ (KZ-spectrum for capillary waves).}
\end{figure}
One can see beginning of formation of the Zakharov-Filonenko spectrum $|a_{\vec k}|^2\sim k^{-17/4}$ (KZ or Kolmogorov-Zakharov spectrum for capillary waves)~\citep{ZakharovPhD,ZF1967,ZLF1992,Nazarenko2011}, corresponding to the constant flux of energy to the small scales (large wavenumbers $k$). At this moment of time average steepness $\mu\approx 0.19$. We see the limited range of scales for the KZ-spectrum, as it still continues to propagate to large $k$'s. What is important, the average steepness is almost the same as an initial one, thus the nonlinear interactions are close to the peak of strength. In the left panel of the next Fig.~\ref{fig:ring5_longer} one can see propagation of the KZ-tail further in larger $k$'s.

\begin{figure}[htb]
\centering
\includegraphics[width=0.45\textwidth]{"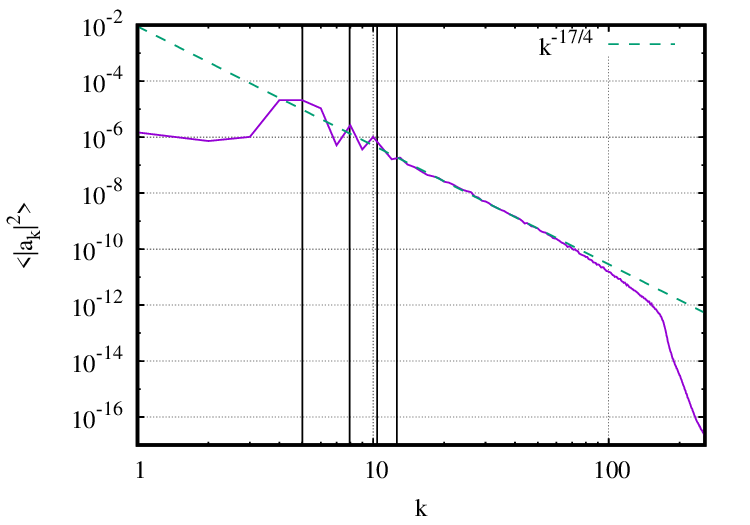"}
\includegraphics[width=0.45\textwidth]{"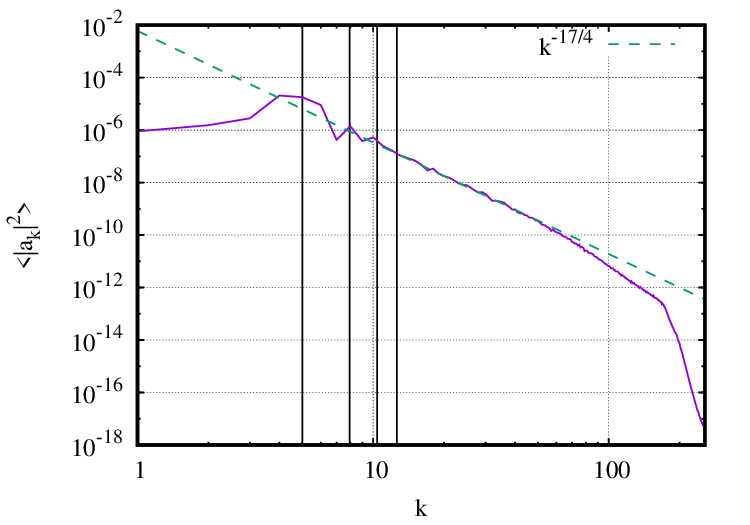"}
\caption{\label{fig:ring5_longer}Angle averaged $|a_{\vec k}|^2$. (Left) Moment of time $t=213 T_{5}$, $\mu\approx 0.19$. (Right) Moment of time $t=404 T_{5}$, $\mu\approx 0.17$. Green line corresponds to weakly turbulent Zakharov-Filonenko spectrum $|a_{\vec k}|^2\sim k^{-17/4}$ (KZ-spectrum for capillary waves).}
\end{figure}
The moment of time $t=213 T_{5}$ corresponds to virtually the same average steepness $\mu\approx 0.19$ as was observed at $t=90 T_{5}$. It is worth to note, that the range of realization of KZ-spectrum for capillary waves in the left panel of Fig.~\ref{fig:ring5_longer} is the widest reported in the literature for now. In the classic work~\citet{PZ1996} the range $[k_{left},k_{right}]$ for clearly visible Zakharov-Filonenko spectrum was $k_{right}/k_{left}\approx 1.8$, in the next series of classic works~\citet{PY2014,PY2015} this range was almost twice wider $k_{right}/k_{left}\approx 3$. Here we see $k_{right}/k_{left}\gtrsim 5$ at least. It should be said that when the second or even third multiple harmonic of the harmonic under investigation is reaching the dissipation region, one have to expect noticeable drain of energy. This mechanism was considered in the case of gravity waves in~\citet{ZKP2009,KPZ2019} and has the same importance for capillary waves. Taking into account that dissipation starts already at $k_d=170$, we should start to see deviation of the spectrum from the one expected in the inertial interval somewhere between $k=60$ (which has third harmonic reaching the dissipation region) and $k=85$ (second harmonic of which is dissipated). 
This range is slightly smaller already at the moment of time $t=404 T_{5}$. It has to be noted that the average steepness at this moment is also smaller $\mu\approx 0.17$, thus nonlinear interactions are weaker, than for $t=213 T_{5}$ (it was shown in~\citet{KPZ2019} that the process of extra dissipation in the formally ``inertial interval'' due to multiple harmonics generation, reaching the dissipation region, is very steep function of average steepness). We will discuss the reason for that in details later. What is important, the visual region of Zakharov-Filonenko spectrum gets smaller, specifically $k_{right}$ is decreasing.

\begin{figure}[htb]
\centering
\includegraphics[width=0.45\textwidth]{"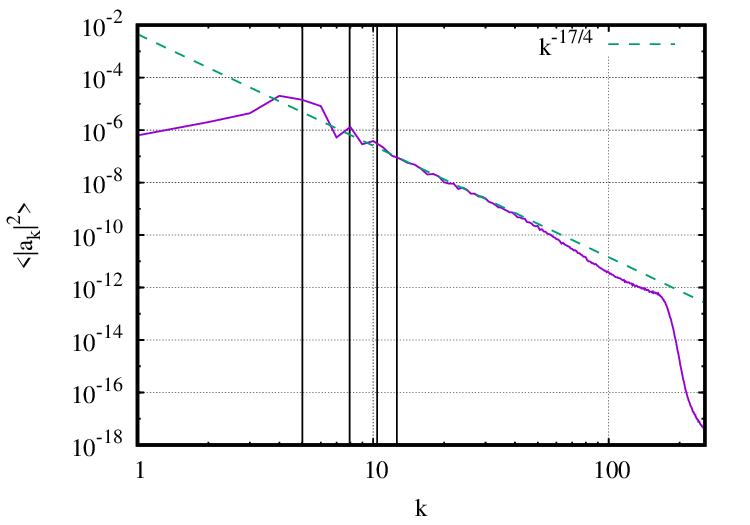"}
\includegraphics[width=0.45\textwidth]{"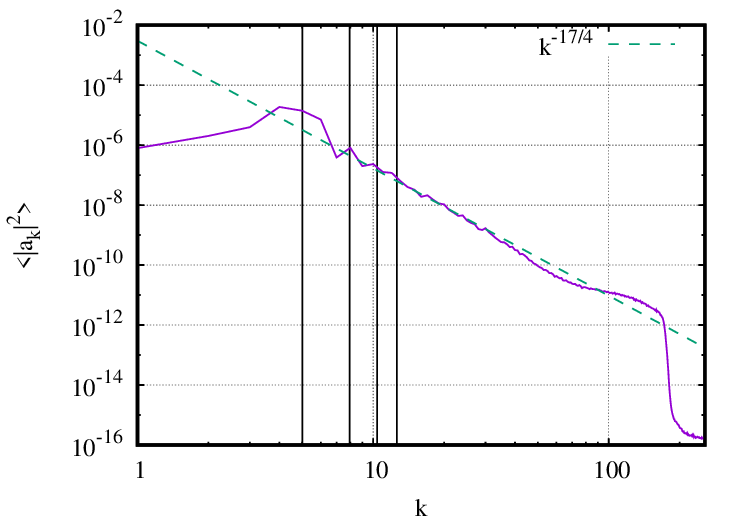"}
\caption{\label{fig:ring5_longest}Angle averaged $|a_{\vec k}|^2$. (Left) Moment of time $t=648 T_{5}$, $\mu\approx 0.15$. (Right) Moment of time $t=1025 T_{5}$, $\mu\approx 0.14$. Green line corresponds to weakly turbulent Zakharov-Filonenko spectrum $|a_{\vec k}|^2\sim k^{-17/4}$ (KZ-spectrum for capillary waves).}
\end{figure}
Further dynamics shows similar processes. Namely, the energy flux continues to take the energy to the dissipation region, average steepness is decreasing from $\mu\approx 0.15$ at $t=648 T_{5}$ to the $\mu\approx 0.14$ at the moment of time $t=1025 T_{5}$. Zakharov-Filonenko spectrum is present, but the region continue to shrink due to decreasing $k_{right}$. Even more, one can see starting from $t=648 T_{5}$ accumulation of waves in the small scale part of the spectrum up to the dissipative scale. In the moment of time $t=1025 T_{5}$ much slower decaying tail is clearly visible.

\begin{figure}[htb]
\centering
\includegraphics[width=0.45\textwidth]{"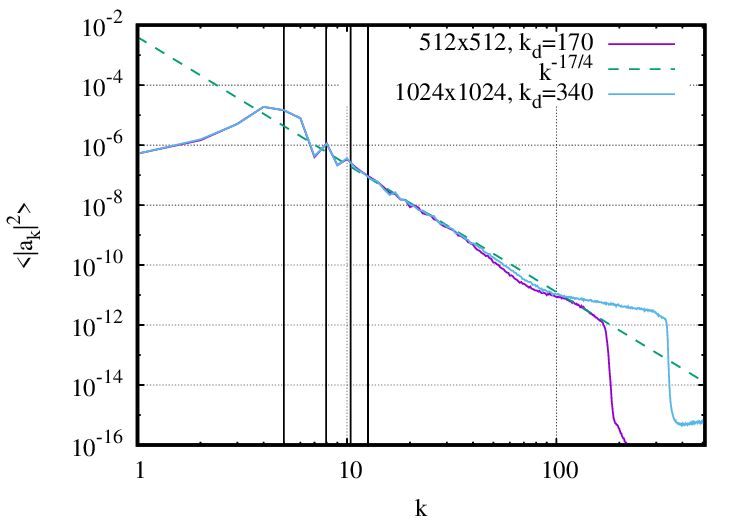"}
\includegraphics[width=0.45\textwidth]{"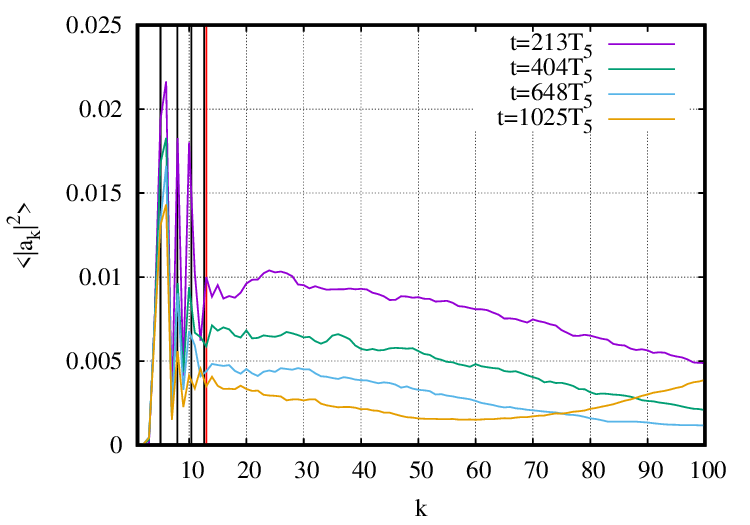"}
\caption{\label{fig:ring5_investigation}Angle averaged $|a_{\vec k}|^2$. (Left) Moment of time $t=802 T_{5}$, two grids $512\times512$ and $1024\times1024$ with dissipative scales $k_d$ equal to $170$ and $340$ correspondingly. Green dashed line corresponds to weakly turbulent Zakharov-Filonenko spectrum $|a_{\vec k}|^2\sim k^{-17/4}$ (KZ-spectrum for capillary waves). (Right) Spectra at different moments of time compensated/multiplied by $k^{17/4}$. To the left of red vertical line simple angle averaged spectrum $|a_{\vec k}|^2$, to the right from it -- the same spectrum after applying moving (sliding) average over the window of 5 harmonics, with an average value in the middle one.}
\end{figure}
One of the obvious reasons for formation of this close to constant tail could be so called ``bottleneck'' phenomenon~\citep{Falkovich1994}, appearing when the dissipation starts too abruptly: accumulation of waves brought by the cascade before the dissipative scale due to absence of the waves in dissipative region to interact with. In order to understand whether this is the only explanation the following experiment was performed: the spectrum from the right panel of Fig.~\ref{fig:ring5_longer} (moment of time $t=404 T_{5}$) was taken and computations were continued on a grid $1024\times1024$ with correspondingly twice further dissipative scale $k_d=340$. The results of this comparison are presented in left panel of Fig.~\ref{fig:ring5_investigation}. As expected, one can see slightly elevated spectrum in small scales due to further boundary of dissipation which results in shift of strongly influenced harmonics (harmonics with third multiple harmonic reaching the dissipation region) from $170/3\approx 57$ to $340/3\approx 113$, but the range of Zakharov-Filonenko spectrum is practically the same. Thus there is another mechanism which prevents continuation of Zakharov-Filonenko spectrum and the ``bottleneck'' is at least not the only mechanism influencing the KZ-spectrum distortion.

\subsection{Possible mechanism of breaking of Zakharov-Filonenko spectrum\label{sec:break_ZF_spectrum}}
Following Appendix~\ref{app:steepness}, if one considers simple model of the isotropic spectrum shown in Fig.~\ref{fig:spectrum_model}:
$$
|a_{k}|^2=
\begin{cases}
A_0^2, k\le k_r,\\
A_0^2 k_r^{17/4} k^{-17/4}, k_r<k<k_d,\\
0, k>k_d.
\end{cases}
$$
where $k_r$ is the radius of the initial ring in $k$-space (see Subsections~\ref{sec:ring43} and~\ref{sec:ring5}) and $k_d$ is the scale from which dissipation starts as explained in Subsection~\ref{sec:num_params}, the relation of the squared average steepness and the parameters of the model spectrum in the case of $k_d\gg k_0$ is the following:
\begin{equation}
\label{C_k_through_mu_sqr}
\mu^2=\langle|\vec\nabla\eta(\vec r)|^2\rangle\approx\frac{L_x L_y}{2\pi}\frac{A_0^2}{\sqrt{\sigma/\rho}}\frac{34}{21}k_r^{7/2},\Rightarrow A_0^2 k_r^{17/4}\approx \frac{2\pi\sqrt{\sigma/\rho}}{L_x L_y}\frac{21}{34}k_r^{3/4}\mu^2.
\end{equation}
The right part of~\eqref{C_k_through_mu_sqr} gives in the inertial interval the following form of Zakharov-Filonenko spectrum:
\begin{equation}
\label{steepness_spectrum}
|a_k|^2 \approx \frac{2\pi\sqrt{\sigma/\rho}}{L_x L_y}\frac{21}{34}k_r^{3/4}\mu^2 k^{-17/4}.
\end{equation}
It has to be said, that the Zakharov-Filonenko spectrum $\sim k^{-17/4}$ corresponding to the constant flux of energy from large to small scales is derived in inertial interval (no injection or dissipation of energy) with assumption, that it is local in $k$-space, meaning determined by interaction of similar scales~\citep{ZakharovPhD,ZLF1992,Nazarenko2011}. Thus, in isotropic case, taking into account where we have maxima of increment (see Fig.~\ref{fig:increment}), defined by maxima of interaction coefficient $M_{\vec k_1 \vec k_2}^{\vec k_0}$ in~\eqref{lambda_merge}, the main process of energy transfer is merging of two waves of close or the same magnitude of wavenumber $k= k_1\simeq k_2$ (radius of the ring in $k$-space) to the third one with a magnitude of the wavenumber $2^{2/3} k\approx 1.6 k$, shown in the left panel of Fig.~\ref{fig:ring43_initial_later}. It is clear that when cascade of energy propagates further into large $k$'s (smaller scales), the $|a_k|^2$ in the increment expression~\eqref{lambda_merge} is decaying as defined in~\eqref{steepness_spectrum}. At the same time time interaction coefficient $M_{\vec k_1 \vec k_2}^{\vec k_0}$ is homogeneous function and is scaling as $M_{\varepsilon \vec k_1, \varepsilon \vec k_2}^{\varepsilon \vec k_0} =\varepsilon^{9/4} M_{\vec k_1 \vec k_2}^{\vec k_0}$, which can be obtained directly from~\eqref{matrix_element}. Now let us estimate how frequency mismatch (which is always nonzero on the evenly spaced grid points in $k$-space, separated by $\Delta k_x=2\pi/L_x$ and $\Delta k_y=2\pi/L_y$) is changing when characteristic scale is getting smaller (larger $k$'s):
\begin{equation}
\label{Delta_Omega_scaling}
\Delta \Omega_{k_1 k_2}^{k_0} \approx \frac{\partial\Omega_{k_1 k_2}^{k_0}}{\partial k}\Delta k,\Rightarrow \Delta \Omega_{\varepsilon k_1, \varepsilon k_2}^{\varepsilon k_0} \approx \varepsilon^{1/2} \Delta \Omega_{k_1 k_2}^{k_0}.
\end{equation}
At the same time the positive (nonlinear) part under the square root sign in~\eqref{lambda_merge} scales as follows:
\begin{equation}
\label{nonlinear_part_scaling}
\left[M_{\varepsilon \vec k_1, \varepsilon \vec k_2}^{\varepsilon \vec k_0}\right]^2 |a_{\vec \varepsilon k}|^2 = \varepsilon^{9/2}\varepsilon^{-17/4} \left[M_{\varepsilon \vec k_1, \varepsilon \vec k_2}^{\varepsilon \vec k_0}\right]^2 |a_{\vec \varepsilon k}|^2 = \varepsilon^{1/4} \left[M_{\varepsilon \vec k_1, \varepsilon \vec k_2}^{\varepsilon \vec k_0}\right]^2 |a_{\vec \varepsilon k}|^2.
\end{equation}
Comparing two scalings~\eqref{Delta_Omega_scaling} and~\eqref{nonlinear_part_scaling} one can see that going further in $k$'s, thus increasing $\varepsilon$ we inevitably will get negative value under square root sign in~\eqref{lambda_merge}, which effectively stops energy transfer using this merging mechanism (when two waves $k = k_1\simeq k_2$ merge into the third wave with wavenumber of magnitude $2^{2/3} k$). Equating approximations for both terms under the square root sign of~\eqref{lambda_merge} and using~\eqref{steepness_spectrum} one can determine scaling of the critical scale $k_c$ where direct cascade stops with respect to key parameters of the simulation:
\begin{equation}
\label{dynamic_range_mu_scaling}
k_c \sim \mu^{8/3} L^{8/3} k_r, 
\end{equation}
where $L$ is the characteristic size of the periodic simulation box in numerical simulations or size of the experimental cell, a value around $L_x$ and $L_y$, which determines $\Delta k=2\pi/L$ in~\eqref{Delta_Omega_scaling}.

Looking at the right panel of Fig.~\ref{fig:ring5_investigation} one can estimate that $k_r\in (7,10)$ (strong peaks corresponding to resonances with initial condition ring, similar to the picture in Subsection~\ref{sec:ring43}, are disturbing the plot) and for $t=213 T_5$ Zakharov-Filonenko spectrum propagates up to the $k_c\in (50,60)$ (this is also confirmed by the left panel of Fig.~\ref{fig:ring5_longer}), while average steepness $\mu\approx 0.19$. At the moment of time $t=1025T_5$ average steepness drops to the value $\mu\approx 0.14$ and according to the right panel of Fig.~\ref{fig:ring5_longest} $k_c\in (20,30)$ (with the upper limit closer to $25$) which is confirmed by the compensated spectrum in the right panel of Fig.~\ref{fig:ring5_investigation}. According to scaling~\eqref{dynamic_range_mu_scaling} we were supposed to get range for $k_c$ approximately $(0.19/0.14)^{8/3}\approx 2.3$ times smaller or $\approx [20,26]$. This is a perfect match taking into account low accuracy of determining of $k_c$. Much better accuracy can be obtained in numerical simulations with pumping, after averaging of the spectrum over time, like in works~\citet{PZ1996,DKZ2004}, because in such a setup it is possible to catch the dynamic equilibrium, when all injected energy is transferred through energy cascade to damping scale and dissipated, giving effectively steady state.

What happens when the we reach the scale where the KZ-cascade stops due to arrest of similar scale interactions? Energy is still brought by the flux, but cannot propagate further, thus we have accumulation of energy at the scale. At the same time, although less efficient due to small values of interaction coefficient, the process of merging interaction with large waves in the strongest ring (initial condition ring in our case or pumping ring in forced turbulence as in works~\cite{PZ1996,PY2014}) continues, as accumulation of energy adds to the sum of squared amplitudes in~\eqref{lambda_merge}. The situation is somewhat similar to mesoscopic turbulence of gravity waves~\citep{ZKPD2005}, when due to discreteness of the simulation grid only few waves are nonlinearly interacting, transferring the flux of conserved quantity and giving qualitatively similar behavior, which is quantitatively different from the WKE description. This is nonlocal process in $k$-space and is not described by the Zakharov-Filonenko solution. Description of such processes can be approached using technique similar to the one used in~\citet{KNPS2024}, which is in future plans of the author.

\section{Conclusion\label{sec:conclusion}}
We performed comprehensive study of weakly nonlinear interactions of capillary waves on the 2D surface of 3D fluid in a setup standard for numerical simulations or laboratory experiments, namely finite size system. Practically every nonlinear interaction was identified with a theoretically predicted one. For longer times it was demonstrated that only contribution of resonant interactions survives as predicted by the theory. After securing these fundamentals, we continued with initial condition geometries typical for simulation of decaying and forced turbulence: azimuth independent amplitudes of harmonics with random phases. It was shown, that in such a setup together with usual same scale interactions, following the locality (in $k$-space) hypothesis of wave turbulence theory, there are also interactions with the strongest waves (e.g. initial condition waves if they are still present) which are nonlocal in Fourier space. Formation of Kolmogorov-Zakharov power-like tails of the spectrum, corresponding to direct energy cascade Zakharov-Filonenko spectrum was observed with a record range of scales around half a decade. Distortion of the spectrum for smaller scales, which is not associated with the artificial dissipation (or at least not only with it), was observed and analytical explanation to it was given. It was demonstrated, that due to increasing influence of discreteness of the wavenumbers grid, which is manifestation of the finiteness of the system, the Zakharov-Filonenko spectrum has a finite range of scales in typical numerical and experimental setups. Scaling of this range of scales with respect to the main parameters of the setup, namely characteristic size of the system and average steepness (measure of nonlinearity) was derived and approximately confirmed for different levels of nonlinearity (values of average steepness). As a byproduct, it was also shown that Kolmogorov-Zakharov spectrum of direct cascade for gravity waves cannot continue infinitely into small scales and has to switch to steeper decaying spectrum, like it is observed in open field experiments. This is a consequence of the fact that average steepness diverges on this spectrum, while the primordial equations are derived in assumption of small average steepness.

This work can be considered as a guide for choosing parameters for numerical or experimental setups, as well as stimulation for future investigations, like accurate confirmation of the proposed scaling in forced turbulence simulations and experiments, which allow more accurate determination of range of scales of realization of Zakharov-Filonenko spectrum. Also analytical theory of nonlocal (in $k$-space) interaction of capillary waves is in the nearest plans of the author.



%
\begin{bmhead}[Acknowledgements]
This work would not be possible without free software from~\citet{GNU},\citet{Gnuplot}, and FFTW library~\citep{FFTW}.
\end{bmhead}
\begin{bmhead}[Funding]
Author would like to thank the Russian Science Foundation for funding this research in the framework of the grant 25-72-31023.
\end{bmhead}
\begin{bmhead}[Declaration of interests]
The author reports no conflict of interest.
\end{bmhead}
\begin{bmhead}[Data availability statement]
The data that support the findings of this study are available from the corresponding author upon reasonable request.
\end{bmhead}
\begin{bmhead}[Author ORCID]
Alexander O. Korotkevich, \url{https://orcid.org/0000-0002-3535-4525}
\end{bmhead}
%

\appendix
\begin{appen}

\section{Relation of the average steepness and isotropic spectrum}\label{app:steepness}
Using the definition of $a_{\vec k}$, $k=|\vec k|$ and using Hermitian symmetry of Fourier components of real function $\eta_{\vec k} = \eta_{-\vec k}^*$ ($^*$ mean complex conjugation) one can express the spectrum of surface elevation:
\begin{equation}
\label{eta_k}
\eta_{\vec k} = \sqrt{\frac{k}{2\omega_k}}(a_{\vec k} + a_{-\vec k}^{*}).
\end{equation}
If one considers Fourier series, which is natural for periodic boundary conditions, in two dimensions ($\vec r = (x,y)$ and $\vec k = (l2\pi/L_x,m2\pi/L_y)$, with $l,m \in \mathbb Z$):
\begin{equation}
f_{\vec k} = \hat F [f(\vec r)] = \frac{1}{L_x L_y}\int\limits_{-L_x/2}^{L_x/2}\int\limits_{-L_y/2}^{L_y/2} f(\vec r) \E^{-\I\vec k\vec r}\D x\D y,\;\;
f(\vec r) = {\hat F}^{-1} [f_{\vec k}] = \sum\limits_{\vec k} f_{\vec k} \E^{\I\vec k\vec r},
\end{equation}
then for the gradient of the surface elevation we have:
\begin{equation}
\label{grad_eta}
\vec\nabla \eta(\vec r) = \sum\limits_{\vec k}\I\vec k \eta_{\vec k} \E^{\I\vec k\vec r}.
\end{equation}
In order to use the following definition of the average steepness:
\begin{equation}
\mu=\sqrt{\langle|\nabla\eta|^2\rangle},
\end{equation}
one needs to apply averaging over the period $L_x\times L_y$ to the square of the~\eqref{grad_eta}:
\begin{align}
\langle|\vec\nabla\eta(\vec r)|^2\rangle=\frac{1}{L_x L_y}\int\limits_{L_x\times L_y}|\vec\nabla\eta(\vec r)|^2\D\vec r &= -\sum\limits_{\vec k}\sum\limits_{\vec k'}(\vec k\cdot\vec k')\eta_{\vec k}\eta_{\vec k'}\frac{1}{L_x L_y}\int\limits_{L_x\times L_y}\E^{\I(\vec k+\vec k')\vec r}\D\vec r\nonumber\\
&=-\sum\limits_{\vec k}\sum\limits_{\vec k'}(\vec k\cdot\vec k')\eta_{\vec k}\eta_{\vec k'}\delta_{\vec k,-\vec k'}=\sum\limits_{\vec k}k^2|\eta_{\vec k}|^2.\label{grad_nabla_2_avrg_Parseval}
\end{align}
This is nothing else by a version of Parseval's identity. Here $\delta_{\vec k,-\vec k'}$ is a standard Kronecker delta. We compute squared magnitude of~\eqref{eta_k}:
\begin{equation}
\label{eta_k_2}
|\eta_{\vec k}|^2 = \frac{k}{2\omega_k}(a_{\vec k} + a_{-\vec k}^{*})(a_{\vec k}^* + a_{-\vec k})=\frac{k}{2\omega_k}(|a_{\vec k}|^2+|a_{-\vec k}|^2 +a_{\vec k}a_{-\vec k} + a_{\vec k}^*a_{-\vec k}^*).
\end{equation}
Finally, substituting~\eqref{eta_k_2} into~\eqref{grad_nabla_2_avrg_Parseval} one get's:
\begin{equation}
\label{mu_sqr_no_radial_sum_general}
\langle|\vec\nabla\eta(\vec r)|^2\rangle=\sum\limits_{\vec k}\frac{k^3}{\omega_k}(|a_{\vec k}|^2+\Re[a_{\vec k}a_{-\vec k}]).
\end{equation}
Up to now everything is exact and symmetry of the spectrum was not yet used.

Let us consider isotropic (independent of an azimuth angle) spectrum, which amplitude depends only on magnitude $k$ of the wavevector $\vec k$ and phases of harmonics are random. If one evaluates term $\sum_{\vec k}(k^3/\omega_k)\Re[a_{\vec k}a_{-\vec k}]$ for some small range $[k,k+\Delta k]$ where amplitudes are almost constants, then this sum over such an interval is proportional to $\langle \cos(\phi)\rangle_{angle}$, where $\phi$ is a uniformly distributed random number $\phi\in[0,2\pi)$. In the case when there are many harmonics in the range $[k,k+\Delta k]$ (equivalent to averaging over many directions), one can use approximation $\langle \cos(\phi)\rangle_{angle}\approx 0$. It is viable to use the following approximation:
\begin{equation}
\label{mu_sqr_radial_sum_general}
\langle|\vec\nabla\eta(\vec r)|^2\rangle\approx\sum\limits_{\vec k}\frac{k^3}{\omega_k}|a_{\vec k}|^2.
\end{equation}
It should be said that isotropic geometry is popular for wave turbulence simulations because every direction in such a case can be considered statistically independent, thus allowing us to average spectrum over the angle as an approximation of averaging over realizations. Obviously, this approximation works the better the more harmonics there are to average over. Meaning that we need to consider relatively large $k$. In the case of averaging over realizations this limitation is not applicable, as $\langle \cos(\phi)\rangle_{ensemble}\rightarrow 0$ with increasing number of realizations, each realization with its own random phase $\phi$ for every product of harmonics $a_{\vec k}a_{-\vec k}$.

In order to go further we need to consider some model of the spectrum. The simple but realistic one is presented in Fig.~\ref{fig:spectrum_model}.
\begin{figure}[htb]
\centering
\includegraphics[width=0.45\textwidth]{"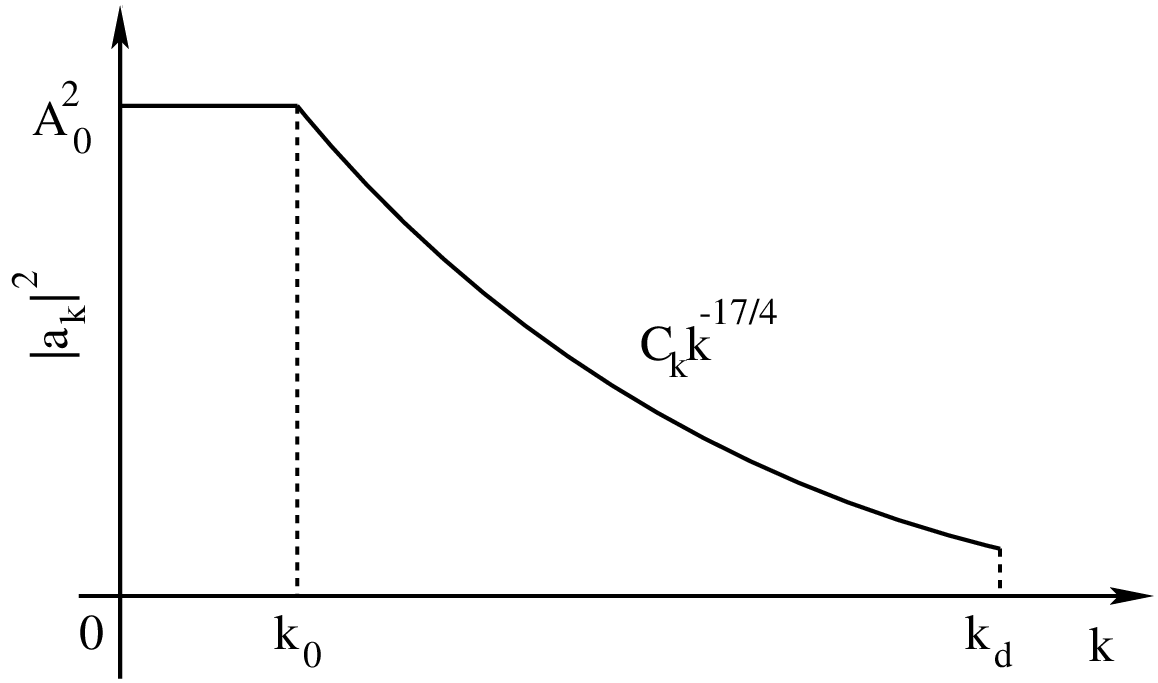"}
\caption{\label{fig:spectrum_model} Model of the spectrum $|a_{k}|^2$. Constant spectrum $A_0^2$ in the interval $k\in[0,k_0]$ which continuously changes into the KZ-spectrum $|a_{k}|^2=C_k k^{-x}$. In this particular figure we considered the case of capillary waves and Zakharov-Filonenko spectrum with $x=17/4$.}
\end{figure}
If we consider works~\cite{PZ1996,PY2014} for capillary waves or papers~\cite{DKZ2003grav,DKZ2003grav,DKZ2004} for gravity waves, one can notice that constant spectrum $|A_0|^2$ up to and a little further the pumping scale can be considered as a rough approximation with attached KZ-spectrum $C_k k^{-x}$ afterwards toward the smaller scales (larger $k$'s). Here we use notation $C_k$ for the constant multiplier in KZ-spectrum following~\cite{PY2014}. In order to evaluate this sum it is reasonable to use $\Delta k_x = 2\pi/L_x$, $\Delta k_y = 2\pi/L_y$ and replace summation by integration in~\eqref{mu_sqr_radial_sum_general}:
\begin{equation}
\label{mu_sqr_radial_int_general}
\langle|\vec\nabla\eta(\vec r)|^2\rangle\approx\frac{L_x L_y}{(2\pi)^2}\sum\limits_{\vec k}\frac{k^3}{\omega_k}|a_{\vec k}|^2\Delta k_x \Delta k_y \approx
\frac{L_x L_y}{(2\pi)^2}\int\limits_0^{+\infty}\frac{k^3}{\omega_k}|a_{k}|^2 2\pi k\D k.
\end{equation}
According to Fig.~\ref{fig:spectrum_model} we model spectrum by the following function:
\begin{equation}
\label{model_n_k_function}
|a_{k}|^2=
\begin{cases}
A_0^2, k\le k_0,\\
A_0^2 k_0^x k^{-x}, k_0<k<k_d,\\
0, k>k_d.
\end{cases}
\end{equation}
As a result of integration of~\eqref{model_n_k_function}, if we put $\omega_k =q k^\alpha$, one gets:
\begin{equation}
\label{mu_sqr_radial_int_model}
\langle|\vec\nabla\eta(\vec r)|^2\rangle\approx\frac{L_x L_y}{2\pi q}\int\limits_0^{+\infty}k^{4-\alpha}|a_{k}|^2 \D k=\frac{L_x L_y}{2\pi}\frac{A_0^2}{q}\left(\left.\frac{k^{5-\alpha}}{5-\alpha}\right|_{0}^{k_0} + k_0^x\left.\frac{k^{5-\alpha-x}}{5-\alpha-x}\right|_{k_0}^{k_d}\right).
\end{equation}

For capillary waves $\omega_k =\sqrt{(\sigma/\rho) k^3}$, where $\sigma$ is surface tension coefficient and $\rho$ is density of the fluid, thus $q=\sqrt{\sigma/\rho}$ and $\alpha=3/2$. KZ-spectrum $\sim k^{-17/4}\Rightarrow x= -17/4$. Which gives after applying all these parameters to~\eqref{mu_sqr_radial_int_model}:
\begin{equation}
\label{mu_sqr_radial_int_model_cap}
\mu^2_{cap}=\langle|\vec\nabla\eta(\vec r)|^2\rangle\approx\frac{L_x L_y}{2\pi}\frac{A_0^2}{\sqrt{\sigma/\rho}}\left(\frac{34}{21}k_0^{7/2} - \frac{4}{3}k_0^{17/4}k_d^{-3/4}\right)=\frac{L_x L_y}{2\pi}\frac{A_0^2}{\sqrt{\sigma/\rho}}\frac{34}{21}k_0^{7/2}\left(1-\frac{14}{17}\frac{k_0^{3/4}}{k_d^{3/4}}\right).
\end{equation}
Taking into account that normally $k_d\gg k_0$, the parenthesis in the last expression can be omitted.

For gravity waves $\omega_k =\sqrt{g k}$, meaning $q=\sqrt{g}$ and $\alpha=1/2$. KZ-spectrum has a slope $\sim k^{-4}\Rightarrow x= -4$. Applying these parameters to~\eqref{mu_sqr_radial_int_model} results in:
\begin{equation}
\label{mu_sqr_radial_int_model_grav}
\mu^2_{grav}=\langle|\vec\nabla\eta(\vec r)|^2\rangle\approx\frac{L_x L_y}{2\pi}\frac{A_0^2}{\sqrt{g}}\left(2k_0^{4}k_d^{1/2} - \frac{16}{9}k_0^{9/2} \right)=\frac{L_x L_y}{2\pi}\frac{A_0^2}{\sqrt{g}}2k_0^{4}k_d^{1/2}\left(1-\frac{16}{18}\frac{k_0^{1/2}}{k_d^{1/2}}\right).
\end{equation}
Once again, if $k_d\gg k_0$, the parenthesis in the last expression might be omitted, although in this case situation is more subtle as we efficiently have a square root of the small parameter.

Comparing these two important cases, one can see that in the case of capillary waves one can take $k_d\rightarrow+\infty$ and this will give just small correction. In other words, average steepness~\eqref{mu_sqr_radial_int_model_cap} is determined by scales from $k=0$ to the somewhat further, but close region around $k=k_0$. Thus, the fact that a constant can be rather crude approximation of the spectrum in the range $k\in[0,k_0]$, can influence the evaluation of the average steepness. While for gravity waves steepness is determined mostly by far tails~\eqref{mu_sqr_radial_int_model_grav}, meaning that behaviour of the spectrum in the beginning of the $k$-axis is not too influential. This was demonstrated in a paper~\cite{KZ2015}, where dependence of $\mu$ on $k_d$ was investigated numerically. Another consequence is the fact, that on a Kolmogorov-Zakharov tail we have divergence of average steepness $\mu$ for $k_d\rightarrow +\infty$. At the same time, our original Hamiltonian dynamical equations are derived as expansion of Hamiltonian in terms of $\mu \ll 1$. This means, that KZ-spectrum for gravity waves cannot go to infinitely small scales, but has to be cut off at some scale by dissipation or switch to a steeper decaying spectrum. This is what usually observed in experiments, where direct cascade ether reach a spectrum similar to a Phillips one
(see, some history of the spectrum in~\citet{Phillips1958,Phillips1967,Kuznetsov2004,NZ2008,Korotkevich2008D} and experimental observation of switching KZ-spectrum for gravity waves to a steeper one in~\cite{Hwang2000}) or to the capillary scale, where it will change to the case~\eqref{mu_sqr_radial_int_model_cap}.

In order to evaluate an average steepness from the wave kinetic equation (WKE) spectrum, one needs to repeat the same computations, taking into account that usually for kinetic equation the symmetric Fourier integral is used, instead of Fourier series:
\begin{equation}
f_{\vec k} = \frac{1}{2\pi}\int\limits_{-\infty}^{+\infty}\int\limits_{-\infty}^{+\infty} f(\vec r) \E^{-\I\vec k\vec r}\D x\D y,\;\;
f(\vec r) = \frac{1}{2\pi}\int\limits_{-\infty}^{+\infty}\int\limits_{-\infty}^{+\infty} f_{\vec k} \E^{\I\vec k\vec r}\D k_x\D k_y.
\end{equation}
The only problem is that there is no reasonable definition of an average steepness for an infinite domain. Even if one would like to apply oceanographical definition $k_p \sqrt{\langle \eta^2\rangle}$, where $k_p$ is a wavenumber of the spectral peak, taking into account that during derivation of weakly nonlinear expansion of the Hamiltonian in terms of steepness we require zero deviation from the unperturbed surface at infinities (see Supplemental Materials for the paper~\cite{Korotkevich2023} for detailed derivation) such a definition also has no sense. So the only approach is to consider a symmetric normalization of Fourier series for a finite periodic region $L_x\times L_y$:
\begin{equation}
f_{\vec k} = \frac{1}{\sqrt{L_x L_y}}\int\limits_{-L_x/2}^{L_x/2}\int\limits_{-L_y/2}^{L_y/2} f(\vec r) \E^{-\I\vec k\vec r}\D x\D y,\;\;
f(\vec r) = \frac{1}{\sqrt{L_x L_y}}\sum\limits_{\vec k} f_{\vec k} \E^{\I\vec k\vec r},
\end{equation}
and then take the limit $L_x,L_y\rightarrow+\infty$ for the result. Here we will reproduce only key points of this straightforward derivation. For the gradient of surface elevation one has:
\begin{equation}
\label{grad_eta_WKE}
\vec\nabla \eta(\vec r) = \frac{1}{\sqrt{L_x L_y}}\sum\limits_{\vec k}\I\vec k \eta_{\vec k} \E^{\I\vec k\vec r}.
\end{equation}
Squared steepness is:
\begin{equation}
\mu^2_{WKE}=\langle|\vec\nabla\eta(\vec r)|^2\rangle=\frac{1}{L_x L_y}\sum\limits_{\vec k}k^2|\eta_{\vec k}|^2=\frac{1}{(2\pi)^2}\sum\limits_{\vec k}k^2|\eta_{\vec k}|^2\Delta k_x\Delta k_y,\label{grad_nabla_2_avrg_Parseval_sum_WKE}
\end{equation}
where $\Delta k_x = 2\pi/L_x$ and $\Delta k_y = 2\pi/L_y$. When we take the limit $L_x,L_y\rightarrow+\infty$ the sum in~\eqref{grad_nabla_2_avrg_Parseval_sum_WKE} turns into the Riemann sum which one can replace with an integral:
\begin{equation}
\mu^2_{WKE}=\langle|\vec\nabla\eta(\vec r)|^2\rangle\xrightarrow[L_x,L_y \to +\infty]{}\frac{1}{(2\pi)^2}\int\limits_{-\infty}^{+\infty}\int\limits_{-\infty}^{+\infty}k^2|\eta_{\vec k}|^2\D k_x \D k_y.\label{grad_nabla_2_avrg_Parseval_WKE}
\end{equation}
This is exactly the limiting transition we were looking for. Applying the same logic as for derivation of~\eqref{mu_sqr_no_radial_sum_general} one gets:
\begin{equation}
\label{mu_sqr_no_radial_int_general_WKE}
\mu^2_{WKE}=\frac{1}{(2\pi)^2}\int\limits_{-\infty}^{+\infty}\frac{k^3}{\omega_k}(|a_{\vec k}|^2+\Re[a_{\vec k}a_{-\vec k}])\D\vec k.
\end{equation}
For isotropic spectrum, expression~\eqref{mu_sqr_no_radial_int_general_WKE} simplifies into:
\begin{equation}
\label{mu_sqr_radial_int_general_WKE}
\mu^2_{WKE}=\frac{1}{2\pi}\int\limits_0^{+\infty}\frac{k^4}{\omega_k}|a_{k}|^2 \D k.
\end{equation}
Taking into account that we use averaging over the angles as replacement to averaging over realizations, one can substitute pair correlator $\langle a_{\vec k}a_{\vec k'}^*\rangle=n_{\vec k}\delta(\vec k - \vec k')$, for which WKE is written, into~\eqref{mu_sqr_radial_int_general_WKE}:
\begin{equation}
\label{mu_sqr_radial_int_general_WKE_n_k}
\mu^2_{WKE}=\frac{1}{(2\pi)^2}\int\limits_{-\infty}^{+\infty}\frac{k^3}{\omega_k}n_{\vec k}\D\vec k=\left\{
\begin{array}{c}
\mathrm{isotropic}\\
\mathrm{case}
\end{array}
\right\}=\frac{1}{2\pi}\int\limits_0^{+\infty}\frac{k^4}{\omega_k}n_k \D k.
\end{equation}
Using spectrum model~\eqref{model_n_k_function}, shown in Fig.~\ref{fig:spectrum_model}, for $n_k$ one gets:
\begin{equation}
\label{mu_sqr_radial_int_model_WKE}
\mu^2_{WKE}=\frac{1}{2\pi}\frac{A_0^2}{q}\left(\left.\frac{k^{5-\alpha}}{5-\alpha}\right|_{0}^{k_0} + k_0^x\left.\frac{k^{5-\alpha-x}}{5-\alpha-x}\right|_{k_0}^{k_d}\right).
\end{equation}
For capillary waves:
\begin{equation}
\label{mu_sqr_radial_int_model_cap_WKE}
\mu^2_{cap, WKE}=\frac{1}{2\pi}\frac{A_0^2}{\sqrt{\sigma/\rho}}\frac{34}{21}k_0^{7/2}\left(1-\frac{14}{17}\frac{k_0^{3/4}}{k_d^{3/4}}\right)\xrightarrow[k_0/k_d \to 0]{} \frac{1}{2\pi}\frac{A_0^2}{\sqrt{\sigma/\rho}}\frac{34}{21}k_0^{7/2}.
\end{equation}
For gravity waves:
\begin{equation}
\label{mu_sqr_radial_int_model_grav_WKE}
\mu^2_{grav, WKE}=\frac{1}{2\pi}\frac{A_0^2}{\sqrt{g}}2k_0^{4}k_d^{1/2}\left(1-\frac{16}{18}\frac{k_0^{1/2}}{k_d^{1/2}}\right)\xrightarrow[k_0/k_d \to 0]{}\frac{1}{2\pi}\frac{A_0^2}{\sqrt{g}}2k_0^{4}k_d^{1/2}.
\end{equation}
\end{appen}\clearpage

\bibliographystyle{jfm}
\bibliography{surfacewaves.bib}


\end{document}